\documentclass[12pt]{article}

\pdfoutput=1 % if your are submitting a pdflatex (i.e. if you have)
\usepackage{slashed}
\usepackage{amsmath,amssymb,graphicx,multicol} 
\usepackage{mathrsfs}
\usepackage{bbold}
\usepackage{epsf,color}
\usepackage{hyperref}
\hypersetup{%
  colorlinks = true,
  linkcolor  = black,
  urlcolor   = blue,
  citecolor  = blue
}
\usepackage{cite}
\usepackage{cancel}
\usepackage{framed}
\usepackage{slashed}
\usepackage{dsfont}
\usepackage[nice]{nicefrac}
\usepackage[utf8]{inputenc}

\newcommand*{\tvec}[1]{\ensuremath{\boldsymbol{\mathrm{#1}}}}           % 3 vec
\newcommand*{\trans}{\mathrm{T}}                     % transposed
\DeclareMathOperator{\re}{Re}
\DeclareMathOperator{\tr}{Tr}
\DeclareMathOperator{\im}{Im}

\newcommand{\nc}{\newcommand}
\nc{\non}{\nonumber}
\nc{\hc}{\hbox {H.c.}} 
\nc{\noi}{\noindent}
\nc{\barx}{\bar{x}}
\nc{\pbarn}{\;\hbox {pb}}
\nc{\fbarn}{\;\hbox {fb}}
\nc{\lsp}{\;\;\;\;\;}
\nc{\Lsp}{\;\;\;\;\;\;\;\;\;\;}  
\nc{\LLsp}{\lspace \lspace}
\nc{\lra}{\longrightarrow}
\nc{\llra}{\longleftrightarrow}
\nc{\beq}{\begin{equation}}  \nc{\eeq}{\end{equation}}
\nc{\bea}{\begin{eqnarray}}  \nc{\eea}{\end{eqnarray}}
\nc{\baa}{\begin{array}}     \nc{\eaa}{\end{array}}
\nc{\bit}{\begin{itemize}}   \nc{\eit}{\end{itemize}}
\nc{\ben}{\begin{enumerate}} \nc{\een}{\end{enumerate}}
\nc{\bce}{\begin{center}}    \nc{\ece}{\end{center}}
\nc{\bpm}{\begin{pmatrix}}   \nc{\epm}{\end{pmatrix}}
\nc{\bvt}{\begin{verbatim}}  \nc{\evt}{\end{verbatim}}
\nc{\vp}{\varphi}
\nc{\hp}{\hat\phi}
\nc{\tp}{\tilde\phi}
\nc{\for}{\;\;{\rm for}\;\;}
\nc{\then}{\;\;{\rm then}\;\;}
\nc{\aaa}{\;\;{\rm and}\;\;}
\nc{\p}{\partial}
\def\gev{~{\rm GeV}}
\def\tb{t_{\beta}}

\def\tanb{\tan\beta}
\def\cotb{\cot\beta}

\def\cosbma{\cos(\beta-\alpha)}

\newcommand{\centeron}[2]{{\setbox0=\hbox{#1}\setbox1=\hbox{#2}\ifdim

\wd1>\wd0\kern.5\wd1\kern-.5\wd0\fi \copy0

\kern-.5\wd0\kern-.5\wd1\copy1\ifdim\wd0>\wd1
                                       \kern.5\wd0\kern-.5\wd1\fi}}
\newcommand{\ltap}{\>\centeron{\raise.35ex\hbox{$<$}}
                               {\lower.65ex\hbox{$\sim$}}\>}
\newcommand{\gtap}{\>\centeron{\raise.35ex\hbox{$>$}}
                               {\lower.65ex\hbox{$\sim$}}\>}

\newcommand{\bal}{\begin{eqalign}}
\newcommand{\eal}{\end{eqalign}}

\newcommand\ZZ{\hbox{\zfont Z\kern-.4emZ}}
\font\zfont = cmss10

\renewcommand{\theequation}{\thesection.\arabic{equation}}
\newcommand{\ra}{\rightarrow}
\newcommand{\eps}{\epsilon}
\newcommand{\dpi}{{\pi_D}}

\begin{document}

\vspace*{1cm}

\begin{center}
{\Large Effective Theories of Dark Mesons
with Custodial Symmetry}
\vspace{1cm}

{Graham D. Kribs$^1$, Adam Martin$^2$, Tom Tong$^1$}

\vspace*{1cm}

$^1$Department of Physics, University of Oregon, Eugene, OR 97403 \\
$^2$Department of Physics, University of Notre Dame, South Bend, IN 46556 \\

\vspace*{1cm}

\begin{abstract} 

Dark mesons are bosonic composites of a new, strongly-coupled 
sector beyond the Standard Model.  
We consider several 
dark sectors with fermions that transform under the electroweak group, 
as arise from a variety of models including 
strongly-coupled theories of 
dark matter (e.g., stealth dark matter), 
bosonic technicolor (strongly-coupled indcued electroweak 
symmetry breaking), vector-like confinement, etc.
We consider theories with two and four flavors under an $SU(N)$ strong group
that acquire variously chiral, vector-like, and mixed contributions
to their masses.  
We construct the non-linear sigma model describing the 
dark pions and match the ultraviolet theory
onto a low energy effective theory that provides the leading interactions
of the lightest dark pions with the Standard Model.
We uncover two distinct classes of effective theories 
that are distinguishable by how the lightest dark pions decay:
``Gaugephilic'': where $\pi^0 \ra Z h$, $\pi^\pm \rightarrow W h$
dominate once kinematically open, and 
``Gaugephobic'': where $\pi^0 \ra \bar{f} f$, $\pi^\pm \rightarrow \bar{f}' f$
dominate.  Custodial $SU(2)$ plays a critical role in determining
the ``philic'' or ``phobic'' nature of a model. 
In dark sectors that preserve custodial $SU(2)$, there is no axial anomaly,  
and so the decay $\pi^0 \ra \gamma\gamma$ is highly suppressed.
In a companion paper, we study dark pion production and decay at 
colliders, obtaining the constraints and sensitivity at the LHC.

\end{abstract}

\end{center}

\newpage

\setlength{\parskip}{0em}
\tableofcontents
\setlength{\parskip}{1em}

\renewcommand{\theequation}{\thesection.\arabic{equation}}
\newcommand{\ns}{\negthinspace}
%%%%%%%%%%%%%%%%%%%%%%%%%%%%%%%%%%%%%%%%%%%%%%%%%%%%%%%%%%%%%%%%%%%%%%%%%%%%%%%%%%%%%%%%%%%%%%%%%%%%%%%%%%%%%%%%%%%%%%%%%%%%%%%%%%%%%%%%%%
\section{Introduction}

We consider extensions of the Standard Model that incorporate
a new, strongly-coupled, confining gauge theory with 
fermion representations that transform under the
electroweak group. 
The notion of a new sector of fields transforming under a new,
strongly-coupled, confining group is a fascinating possibility
for physics the Standard Model.  All of the new sector's scales 
are either natural (the new confinement scale) or technically 
natural (new fermion masses), and so such a scenario is,
at a minimum, no worse off than the Standard Model from a 
naturalness point of view.

There are a wide variety of uses of a new, strongly-coupled, 
confining group.  One use is to at least partially break 
electroweak symmetry dynamically, such as bosonic technicolor~\cite{Simmons:1988fu,Samuel:1990dq,Dine:1990jd,Kagan:1990az,Kagan:1991gh,Carone:1992rh,Carone:1993xc,Dobrescu:1997kt,Antola:2009wq} 
and the closely related ideas on strongly-coupled
induced electroweak symmetry breaking~\cite{Azatov:2011ht,Azatov:2011ps, Gherghetta:2011na,Galloway:2013dma,Chang:2014ida,Beauchesne:2015lva,Harnik:2016koz,Alanne:2016rpe,Galloway:2016fuo,Agugliaro:2016clv,Barducci:2018yer}. 
Composite Higgs theories also posit a new strongly-coupled sector 
in which at least an entire Higgs doublet emerges in the 
low energy effective theory (the literature is far too vast to 
survey, for a review see e.g., \cite{Bellazzini:2014yua}). 
There is also a interesting connection to the
relaxation of the electroweak scale \cite{Graham:2015cka}
using a new strongly-coupled sector, e.g., 
\cite{Graham:2015cka,Antipin:2015jia,Agugliaro:2016clv,Batell:2017kho,Barducci:2018yer}.

Dark matter can emerge as a composite meson or baryon of a
strongly-coupled theory, often with an automatic accidental symmetry 
that protects against its decay. 
Since the early days of technicolor there was a possibility of 
dark matter emerging as technibaryons
\cite{Nussinov:1985xr,Chivukula:1989qb,Barr:1990ca,Barr:1991qn,Kaplan:1991ah,Chivukula:1992pn,Bagnasco:1993st}.  
There is now a growing literature
that has studied strongly-coupled dark matter as dark pions
\cite{Khlopov:2008ty,Ryttov:2008xe,Hambye:2009fg,Bai:2010qg,Lewis:2011zb,Frigerio:2012uc,Buckley:2012ky,Bhattacharya:2013kma,Hietanen:2014xca,Marzocca:2014msa,Pasechnik:2014ida,Antipin:2014qva,Hochberg:2014kqa,Carmona:2015haa,Lee:2015gsa,Hochberg:2015vrg,Bruggisser:2016ixa,Ma:2017vzm,Davoudiasl:2017zws,Berlin:2018tvf,Choi:2018iit,Hochberg:2018rjs}, 
dark quarkonia-like states
\cite{Alves:2009nf,Kribs:2009fy,Lisanti:2009am,Alves:2010dd,Geller:2018biy},
as well as dark baryons and related candidates 
\cite{Gudnason:2006yj,Dietrich:2006cm,Foadi:2008qv,Khlopov:2008ty,Mardon:2009gw,Kribs:2009fy,Sannino:2009za,Barbieri:2010mn,Belyaev:2010kp,Lewis:2011zb,Appelquist:2013ms,Hietanen:2013fya,Cline:2013zca,Appelquist:2014jch,Krnjaic:2014xza,Hietanen:2014xca,Detmold:2014qqa,Detmold:2014kba,Brod:2014loa,Asano:2014wra,Appelquist:2015yfa,Appelquist:2015zfa,Drach:2015epq,Fichet:2016clq,Co:2016akw,Dienes:2016vei,Ishida:2016fbp,Francis:2016bzf,Lonsdale:2017mzg,Berryman:2017twh,Mitridate:2017oky,Francis:2018xjd} (for a review, see \cite{Kribs:2016cew}).

Another use is to simply characterize generic strongly-coupled-like 
signals as targets for LHC and future colliders. 
Vector-like confinement \cite{Kilic:2009mi} pioneered this study
in the context of vector-like fermions that transform under
part of the SM group as well as under a new, strongly-coupled group
with scales near or above the electroweak scale.
Further explorations into the phenomenology and especially the
meson sector included
\cite{Kribs:2009fy,Kilic:2010et,Harnik:2011mv,Fok:2011yc,Buckley:2012ky,Bai:2013xga,Brod:2014loa,Chacko:2015fbc,Agashe:2016rle,Matsuzaki:2017bpp,Barducci:2018yer,Draper:2018tmh,Buttazzo:2018qqp}.  
In theories with somewhat lower confinement scales,
the dark sector may lead to dark showers and related
phenomena~\cite{Schwaller:2015gea,Cohen:2015toa,Freytsis:2016dgf,Zhang:2016sll,Cohen:2017pzm,Beauchesne:2017yhh,Renner:2018fhh},
displaced signals \cite{Mahbubani:2017gjh,Buchmueller:2017uqu}
and potentially intriguing spectroscopy
\cite{Hochberg:2015vrg,Daci:2015hca,Hochberg:2017khi}.
Spectacular ``quirky'' signals can arise in theories 
with a very low confinement scale \cite{Han:2007ae,Kang:2008ea}. 
The latter theories may also lead to a high multiplicity of
soft particles that are tricky to observe
\cite{Harnik:2008ax,Knapen:2016hky,Pierce:2017taw}.

The dark sector theory that is of particular interest to us 
is Stealth Dark Matter \cite{Appelquist:2015yfa}.
In this theory, there is a new, strongly-coupled ``dark sector''
that consists of vector-like fermions that transform under both the 
new ``dark group'' group as well as the electroweak part of the SM, 
and crucially, also permit Higgs interactions.  
Others have also pursued dark sectors with vector-like fermions
that permit Higgs interactions for a variety of purposes \cite{Antipin:2014qva,Antipin:2015jia,Beylin:2016kga,
Mitridate:2017oky,Barducci:2018yer}.
The meson sector of the Stealth Dark Matter theory, 
however, has several intriguing properties due to the 
accidental symmetries of the model.

One might think a dark meson sector whose low energy effective theory
is a set of scalars with electroweak quantum numbers has already been 
fully (or mostly) covered by the wide range of existing 
search strategies. 
As we show in our companion paper~\cite{Kribs:2018ilo}, this is \emph{not}
the case.  There we find that a dark vector meson could be
as light as about $300$~GeV, something that, at first glance, 
seems hard to believe given the multi-TeV bounds on new 
$Z'$ bosons from LHC data.  The dark vector meson can mediate
dark pion pair production 
(just like $\rho \ra \pi\pi$ in QCD), and in some models, 
the bounds on the dark pion mass could be as small as $130$~GeV\@.
Clearly, the LHC easily has the energy to produce these states,
and so it really comes down to finding search strategies
that maximize sensitivity.  We believe substantial improvements 
are possible, providing impetus and breathing new life into 
LHC searches in the hundreds of GeV regime.

The difficulty with strongly-coupled physics is that
it is strongly-coupled.  However, many years ago Kilic, Okui, 
and Sundrum pioneered the study of a new strongly-coupled sector's
phenomenology for collider physics~\cite{Kilic:2009mi}.  
Their insight was to determine the leading interactions of 
an effective theory of pseudo-Nambu Goldstone (pNGB) mesons with
vector mesons (both composite and fundamental).  They were motivated 
by imagining QCD scaled up to weak scale energies, except, and here is the
key point, their BSM fermion masses were taken to be 
purely vector-like.

In this paper, we generalize vector-like confinement by permitting 
specific interactions between the strong sector fermions and the 
Standard Model.  In some models, these interactions are 
renormalizable Yukawa couplings of the dark sector fermions with 
the Higgs of the SM\@.  In others that do not permit Yukawa couplings, 
we also consider higher dimensional operators (that also involve
the Higgs sector in some way). 
These interactions lead to dark pion decay.
And, what is distinct in the vector-like theories we consider
is that there is no axial anomaly contribution to neutral 
dark pion decay.
We use a non-linear sigma model (NLSM) to describe
the pNGB mesons, which we carry out in detail in the paper. 
Equally important, the fact that we break the flavor symmetries of 
the strong sector with Higgs interactions necessarily locks the 
strong sector flavor symmetries to the 
$O(4) \cong SU(2)_L \times SU(2)_R$ global symmetry of the 
Higgs potential. As a result, the strong sector fields can be 
grouped into multiplets of this symmetry, with different assignments 
possessing qualitatively different phenomenology.
 
The structure of this paper is as follows. First (Sec.~\ref{strong}), 
we briefly remind the reader of the ingredients in the type of 
strong sector we want to consider.  Next, in 
Sec.~\ref{sec:custodial} we discuss custodial $SU(2)$ of the Higgs sector,
emphasizing the role of hypercharge and the difference
between up-type and down-type fermion Yukawa couplings
that act as the spurions for custodial $SU(2)$ violation.
This will greatly assist us in understanding and classifying 
the dynamics of dark mesons in the set of theories we 
consider.  In Sec.~\ref{sec:twoflavor}, we discuss two-flavor theories, 
one chiral and two vector-like scenarios. 
Understanding the dynamics of these relatively simple 
theories provides a warmup to theories with more flavors.
Next we consider vector-like four-flavor theories, that are the 
smallest field content that permit vector-like masses
and Higgs interactions at the renormalizable level.
The model was first proposed in \cite{Appelquist:2015yfa,Appelquist:2015zfa}
where baryonic sector of these theories was extensively studied
since the lightest baryon is a viable dark matter candidate.
Our main goal is to determine the dark pion interactions
with the SM, and to understand the results in terms of limits
when two of the flavors are decoupled and the theory reduces to
just a two-flavor theory with higher dimensional interactions.
Finally, to emphasize the role of custodial $SU(2)$, 
we discuss a vector-like two-flavor theories where 
custodial symmetry is violated, and the consequences
for the dark pion decays.  In Appendix~\ref{app:2HDM},
we review the case of a general two-Higgs doublet model
and the ``gaugephobic'' decays of its $A^0, H^\pm$ states.

\section{Defining the dark sector}
\label{strong}

Throughout this paper, we will refer to the new strong sector as the 
``dark'' sector. It consists of a strongly-coupled ``dark gauge group'' 
$SU(N_D)$ with its own ``dark confinement scale'', and  ``dark fermions'' 
or ``dark flavors'' that transform under the dark group 
as well as the electroweak part of the Standard 
Model. Below the dark confinement scale, the effective theory
description of the composites includes ``dark baryons''
and ``dark mesons''; the latter breaking up into 
``dark vector mesons'' and ``dark pions''.  
Despite the naming convention, we emphasize that the new states 
are certainly not ``dark'' to collider experiments 
\cite{Kribs:2018ilo}.

When describing the fermionic content of the dark sector (in the UV), we will work entirely with left-handed fields, meaning (1/2, 0) under the Lorentz group. We will distinguish between theories by the number of dark fermion flavors, where each flavor corresponds to one (two-component) fermion in the fundamental of the dark color group and one anti-fundamental.  We will generically refer to dark color fundamentals as $F$, and anti-fundamental as $\hat F$. Throughout this paper, we will assume that all dark fermions are inert under SM $SU(3)_c$ while at least some of them interact electroweakly.  Other references
that have pursued dark sectors transforming under $SU(3)_c$ 
can be found in~\cite{Kilic:2009mi,Bai:2010qg,Bai:2010mn}.  

In the absence of other interactions, the symmetry of the dark sector is $SU(N_{fund})\times  SU(N_{anti})$.  Turning on electroweak interactions, some of these flavor symmetries are explicitly broken.  The majority of the dark sectors we'll study are vector-like, which -- in terms of two-component fermions -- implies that if $F_{i}$ is a fundamental of dark color and transforms under EW representation $G$, then the theory also includes a dark-color anti-fundamental $\hat F_{j}$ also residing in EW representation $G$. This charge assignment permits mass terms of the form $M_{ij} F_{i}\hat F_{j}$. In addition, we can form dimension~$>3$ operators connecting dark fermions with the Higgs boson. Interactions with the Higgs force us to connect flavor symmetries of the fermionic sector with the $O(4) \cong SU(2)_L \times SU(2)_R$ global symmetry Higgs potential. If $F$ are EW doublets and $\hat F$ are EW singlets, then the interactions take a form familiar from SM Yukawas,  $y F\, \hat F\, \mathcal{H}$. For other representations of $F,\hat F$, the interactions only come about at the non-renormalizable level, e.g., $F\, \hat F\, H^{\dag}H/\Lambda$. 

Once we cross below the dark confining scale, the low energy effective theory is described in terms of the composite mesons and baryons of this sector. 
 Provided that the vector-like dark fermion masses are $< 4\pi\, f$, the leading interactions of the dark pions can be determined using non-linear sigma model language analogous to the real pions of QCD. Confinement spontaneously breaks the chiral symmetry of the dark fermions down to the diagonal subgroup: $SU(N_{fund}) \times SU(N_{anti}) \to SU(N)_V$, with the dark pion multiplets falling into representations of $SU(N)_V$. Whether or nor $SU(N)_V$ is gauged and how it connects with the Higgs potential symmetries depends on the setup.  In the IR, interactions between dark fermions and Higgses become interactions between the dark pions and the Higgs. For example, the two examples used above become $tr(\Sigma \mathcal{H}^\dagger) + h.c.$ and ${\rm Tr}(\Sigma \mathcal{H}^{\dag} \mathcal{H} + h.c.)$ respectively, where $\Sigma$ is the NLSM field.
  
At this point it is useful to distinguish between 
the dark sectors that we consider in this paper
and early proposals for dynamical electroweak
symmetry breaking (technicolor).   Simply put, 
in the extension we consider, we \emph{assume} there is a 
Higgs doublet in the low energy effective theory that acquires 
an electroweak breaking vev that is responsible for (most) of
electroweak symmetry breaking in the Standard Model.

\section{Custodial SU(2)}
\label{sec:custodial}

A critical part of the classification of effective theories
of dark mesons is whether custodial $SU(2)$ is preserved or
violated by the dark sector dynamics.  Custodial $SU(2)$
is the residual accidental global symmetry of the 
Higgs multiplet after it acquires an expectation value,
$O(4)\cong SU(2)_L \times SU(2)_R  \ra O(3) \cong SU(2)_C$.  

Custodial $SU(2)$ arises automatically once the matter content 
and interactions are (at least formally) promoted to become 
$SU(2)_L \times SU(2)_R$ invariant. We will use the terminology $SU(2)_L$ , $SU(2)_R$ frequently in this paper and emphasize that this will always refer to internal symmetries of the theory and never to Lorentz symmetry. It will become very 
convenient to utilize a manifestly $SU(2)_C$ symmetric 
formalism for writing interactions of the dark sector with the 
Higgs multiplet.  The basic notions are well-known, though not 
necessarily exploited in the ways that we will be doing.
A manifestly custodially $SU(2)_C$ symmetric formalism
promotes $U(1)_Y$ to $SU(2)_R$, where only the $t_3$
generator of $SU(2)_R$ is gauged.  

To establish notation, the Higgs doublet of the Standard Model
\begin{eqnarray}
H &=& \left( \begin{array}{c} 
             G^+ \\
             (v + h + i G^0)/\sqrt{2} 
             \end{array} \right) \, ,
\end{eqnarray}
can be re-expressed in terms of a $(\mathbf{2},\mathbf{2})$ 
bifundamental scalar field under $SU(2)_L \times SU(2)_R$ as
\begin{eqnarray}
\mathcal{H}_{i_L i_R} &=& \frac{1}{\sqrt{2}}
   \left( \begin{array}{cc}
   (v + h - i G^0)/\sqrt{2} & G^+ \\
   - G^-                    & (v + h + i G^0)/\sqrt{2} 
   \end{array} \right) \, .
\label{eq:h2x2}
\end{eqnarray}
In principle, all custodially-symmetric interactions can be 
written in terms of powers of $\mathcal{H}$, and suitable $SU(2)_L$
and $SU(2)_R$ contractions. The notation becomes much more compact when we utilize
the definition
\begin{equation}
\mathcal{H}^\dagger_{i_R i_L} \, \equiv \, 
\epsilon_{i_R j_R} \epsilon_{i_L j_L} \mathcal{H}_{i_L i_R} 
\end{equation}
which matches the naive complex conjugation and transpose of 
the $2 \times 2$ matrix definition in Eq.~(\ref{eq:h2x2}).
In this form, the Standard Model Higgs potential becomes simply
\begin{eqnarray}
V &=& m_H^2 \mathrm{Tr} \, \mathcal{H}^\dagger \mathcal{H}
     + \frac{\lambda}{4} 
 \left( \mathrm{Tr} \, \mathcal{H}^\dagger \mathcal{H} \right)^2 \, .
\end{eqnarray}
The absence of any explicit $t^3_R$ signals the absence of
any explicit custodial symmetry violation.  When the Higgs gets a vev 
and $SU(2)_L \times SU(2)_R$ breaks to the diagonal $SU(2)_C$, 
the original ({\bf 2},{\bf 2}) of Higgs states decomposes into a 
singlet (radial mode) plus a triplet (Goldstones) of the 
diagonal $SU(2)_C$.

The full covariant derivative for the Higgs multiplet, 
Eq.~(\ref{eq:h2x2}),  does not respect
$SU(2)_R$ due to gauging hypercharge, i.e., just the $t^3_R$ generator
is gauged.  This is straightforwardly handled by writing
the covariant derivative as 
\begin{equation}
D_\mu \mathcal{H}_{i_L i_R} \, = \,
\partial_\mu \mathcal{H}_{i_L i_R} 
-i g W^a_\mu (t^a_L \mathcal{H})_{i_L i_R} 
-i g' B_\mu (\mathcal{H} t^3_R)_{i_L i_R} 
\label{eq:higgs2x2covar}
\end{equation}
making kinetic term of the bi-doublet $\mathcal{H}$:
\begin{eqnarray}
\mathrm{Tr} \, D_\mu \mathcal{H}^\dagger D^\mu \mathcal{H} \, .
\end{eqnarray}
The explicit $t^3_R$ can be thought of as
$2 Y t^3_R$, where the Higgs doublet $H_{Y=1/2}$ 
and its complex conjugate $H^*_{Y=-1/2}$ are embedded
as the two components of an $SU(2)_R$ doublet. 
In the limit $g' Y \rightarrow 0$, the last term of
Eq.~\eqref{eq:higgs2x2covar} vanishes, restoring
the full $SU(2)_R$ global symmetry.  
In this way, we see that $g' Y t^3_R$ acts as a spurion
for custodial $SU(2)$ violation.  
One could instead promote 
$B_\mu t^3_R \rightarrow W_R^a t^a_R$,
formally gauging the full $SU(2)_R$ symmetry. 
In this case, we would need  an  
explicit $SU(2)_R$-breaking mass term in order to remove the 
$W_R^{1,2}$ gauge bosons and recover the Standard Model.  
Moreover, as is well-known from left-right models,
an additional $U(1)$ is required to obtain the correct 
hypercharge of the left-handed and right-handed quarks
and leptons (e.g., for a review, see \cite{Duka:1999uc}). 

Yukawa couplings are another source of custodial breaking. 
In terms of the usual Higgs doublet $H$, the up and down
Yukawa couplings are
\begin{eqnarray}
y^u_{ij} {Q_L}_i \epsilon H {u_R}_j + 
y^d_{ij} {Q_L}_i H^\dagger {d_R}_j  + h.c.
\end{eqnarray}
Grouping ${Q_R}_i = \{ u^c_i, d^c_i\}$ together, we can rewrite the 
up and down quark Yukawas in terms of $\mathcal{H}$ as
\begin{eqnarray}
y^{u}_{ij} {Q_L}_i \mathcal{H} \, P_u {Q_R}_{j} + 
y^{d}_{ij} {Q_L}_i \mathcal{H} \, P_d {Q_R}_{j} + h.c. 
\label{eq:yuk1}
\end{eqnarray}
where $P_{u,d} = (\mathds{1}_R \mp 2 t^3_R)/2$ are matrices in $SU(2)_R$ space
that project out the up-type or down-type right-handed fermion.
In fact, it is useful to rewrite Eq.~\eqref{eq:yuk1} as the sum 
of a custodial symmetric Yukawa plus a custodial violating term:
\begin{eqnarray}
\mathscr{L}_{\rm Yuk} &=& 
  \mathcal{Y}^C_{ij} 
  {Q_L}_i \mathcal{H} \frac{\mathds{1}_R}{2} {Q_R}_j +
  \mathcal{Y}^{\slashed{C}}_{ij} 
  {Q_L}_i \mathcal{H} t^3_R {Q_R}_j  + h.c. 
\label{eq:yuk2}
\end{eqnarray}
where
\begin{equation}
\begin{array}{rcl}
  \mathcal{Y}^C_{ij}               &=& y^u_{ij} + y^d_{ij} \\ 
  \mathcal{Y}^{\slashed{C}}_{ij} &=& y^u_{ij} - y^d_{ij} \, . 
\end{array}
\label{eq:Yeq}
\end{equation}
There is no loss of generality from the SM, i.e., 
$\mathcal{Y}^C_{ij}$ and $\mathcal{Y}^{\slashed{C}}_{ij}$
are independent matrices.  In the special case where 
$y_{ij}^u = y_{ij}^d$ and thus $\mathcal{Y}^{\slashed{C}}_{ij} = 0$,
the Yukawa couplings are custodially symmetric.
Later in the paper when we write higher dimensional operators
involving the SM fermions, we will always assume a form of 
minimal flavor-violation (MFV) where operators involving 
${Q_L}_i \mathcal{H} (\mathds{1}_R/2) {Q_R}_j$ are accompanied 
by $\mathcal{Y}^C_{ij}$ and operators involving 
${Q_L}_i \mathcal{H} t_R^3 {Q_R}_j$ are accompanied by 
$\mathcal{Y}^{\slashed{C}}_{ij}$.

Looking beyond the SM, we will use the same logic we applied to SM Yukawas when writing down interactions between the dark fermions and the Higgs. Specifically, in addition to grouping dark fermions into multiplets of (gauged) $SU(2)_W \equiv SU(2)_L$, we also assign them to multiplets of $SU(2)_R$ then classify interactions in $SU(2)_L \times SU(2)_R$ language. Put another way, interactions among the SM Higgs multiplet and dark fermions break the combination of the $SU(2)_R$ Higgs potential symmetry and the $SU(N_{fund})$ (or $SU(N_{anti})$) flavor symmetries of the dark fermions down to a common $SU(2)$, which we relabel as $SU(2)_R$. New custodial violating breaking interactions/spurions must be proportional to $t^3_R$, as that is the only choice consistent with gauging $SU(2)_L$ and the $t^3$ generator 
[$U(1)_Y$] of $SU(2)_R$~\cite{Buchalla:2014eca}. Thus, in $SU(2)_L \times SU(2)_R$ language, strong sectors that respect custodial symmetry contain no terms with explicit $t^3_R$, while a generic  custodial violating dark sector can have one or more such terms.\footnote{Additionally, dark sector theories
with $SU(2)_L$ multiplets with hypercharge, as well as
$SU(2)_R$ multiplets with hypercharge not proportional to
$t^3_R$, require an additional $U(1)_X$.  We do not consider 
such theories in this paper.}

\section{Effective Interactions of Dark Pions}

The dark sectors of greatest interest to us in this paper 
preserve custodial $SU(2)$, so all deviations from exact custodial symmetry 
can be traced to $g' Y$ or the differences among SM Yukawas. Consequently, 
dark pions transform in representations of $SU(2)_L \times SU(2)_R$.  Once the Higgs gets a vacuum expectation, these pions will break up into multiplets of custodial $SU(2)$. The smallest, and therefore lightest, non-trivial $SU(2)_C$ representation the pions can fill is the triplet. Heavier dark pions in larger representations are possible, as are higher spin composites such spin-1 dark rho mesons.  In general, these states rapidly decay into the lightest dark pions.  While this is certainly highly relevant for phenomenology \cite{Kilic:2009mi,Kribs:2018ilo}, it the lightest dark pion decays that
are the main concern for this paper.  

\subsection{Dark Pion Triplet interactions in custodial preserving strong sectors}
 \label{sec:triplets}

Suppose we have an $SU(2)_C$ triplet of dark pions $\pi^a$, 
that we have already motivated as arising in a wide class of 
interesting class of dark sector theories, 
and we wish to understand its interactions. The most phenomenologically relevant interactions to determine are those with a single dark pion since they will govern decays. As we will show below, single pion interactions can be understood from symmetry considerations alone.

First, let's consider a ``toy'' Standard Model that is fully 
$SU(2)_L \times SU(2)_R$ symmetric -- meaning we set $g' = 0$ and $\mathcal{Y}^{\slashed{C}}_{ij} = 0$, in the presence of a dark sector
that produces a (custodially symmetric) triplet of dark pions.
In this limit, the $\{u^c, d^c\}$ quarks of the SM can be written in terms of  a $SU(2)_R$ doublet as in Eq.~\eqref{eq:yuk1} and the $SU(2)_W$ gauge bosons lie in a $SU(2)_L$ triplet. When EWSB occurs, $SU(2)_L \times SU(2)_R \to SU(2)_C$, so we can reclassify all fields into $SU(2)_C$ multiplets and form invariants from them. Contracting $\pi^a$ with $SU(2)_C$ triplets formed from SM fields,  the lowest dimension operators involving  a single dark pion are:
 \begin{eqnarray}
\mathcal{Y}^C_{ij} \, \Big(\frac{v}{v_\pi}\Big) \, \pi^a 
   \left( {Q_L}_i t^a {Q_R}_j \right) 
   + \xi \, g \, \Big(\frac{v}{v_\pi}\Big) \, W^a_\mu 
     \left( h \overleftrightarrow \partial^\mu \pi^a \right) \, , 
\label{eq:sm-pure-custodial}
\end{eqnarray}
where $t^a$ are the generators of $SU(2)_C$.  
(A similar expression for the first term is also present for 
the leptons of the SM.) 
As both terms require electroweak symmetry breaking, they must be proportional to the mass of the SM fields.  Therefore, we need another dimensionful parameter $v_{\pi}$ to balance dimensions.  For the fermion terms, we have assumed the flavor structure obeys minimal flavor violation with a
(lowest order) coefficient of $\mathcal{Y}^C_{ij}$.
The factor $\xi$ parameterizes the relative strength between the interactions of pions with fermions versus the gauge/Higgs sector. We will explore the size and origin of $v_{\pi}$ and $\xi$ in specific theories shortly. The presence of the Higgs boson in the second term is also easy to motivate.  While $W^a_{\mu} \partial_\mu \pi^a$ is custodially symmetric, by itself this is a mixing involving longitudinal $W^a$ and would indicate that we have not properly gauge fixed.  Hence, we need to add a $SU(2)_C$ singlet, and $h$ is the option with the lowest dimension in the broken phase of the SM. 

One might wonder how $\xi$ could be different from unity,
given what we have described thus far.  
When dark pions transform in representations that are larger
than a triplet, there is a possibility of dark pion--Higgs boson
mixing.  For example, dark pions in the complex representation 
$({\bf 2},{\bf 2})$ under 
$SU(2)_L \times SU(2)_R$ contain a  ``Higgs-like'' 
dark pion state ($SU(2)_C$ singlet) that 
can -- and generically does -- mix with the SM ($SU(2)_C$ singlet) Higgs boson. 
This implies additional contributions to the gauge/Higgs boson/dark pion
interactions arise from the covariant derivative of the dark pions.
These interactions turn out to be \emph{critical} to understanding
the phenomenology of models with more than two flavors of dark fermions.

Let's now re-introduce the custodial $SU(2)$ violation in the SM\@.
This involves the difference between up and down Yukawas, 
\begin{eqnarray}
\mathcal{Y}^{\slashed{C}}_{ij} 
\left( \frac{v}{v_\pi} \right) \, 
\pi^a \left( {Q_L}_i t^a t^3_R {Q_R}_j \right) 
\end{eqnarray}
as well as $g' \not= 0$, 
\begin{eqnarray}
\xi \, g' \, \Big(\frac{v}{v_\pi}\Big) \, B_\mu 
     \left( h \overleftrightarrow \partial^\mu \pi^0 \right) \, .
\end{eqnarray}
With these terms, the simple lagrangian
Eq.~\eqref{eq:sm-pure-custodial} becomes somewhat more complicated.
If we focus our attention on just one generation of quarks, 
and convert from two-component fermions to four-component notation, 
the effective lagrangian for dark pion decay becomes:
\begin{eqnarray} 
\mathscr{L}_{\rm decay} &=& \frac{\sqrt 2}{v_\pi} \, 
\bigg[ \, \dpi^+ \bar{\psi}_u (m_d P_R - m_u P_L) \psi_d \, 
+ \, \dpi^- \bar{\psi}_d (m_d P_L - m_u P_R) \psi_u \nonumber \\
& &{} \qquad\qquad\quad + \frac{i}{\sqrt{2}} \, \dpi^0 (m_u \, \bar{\psi}_u \gamma_5 \psi_u - m_d \, \bar{\psi}_d \gamma_5 \psi_d) \, \bigg]  \nonumber \\
& & - \, \xi \, \frac{m_W}{v_\pi} 
\left[ (W_\mu^-\, h\overleftrightarrow \partial^\mu \dpi^+) + (W_\mu^+\, h \overleftrightarrow \partial^\mu \dpi^-) 
+ \frac{1}{\cos{\theta_W}} (Z_\mu\, h \overleftrightarrow \partial^\mu \dpi^0) \right] 
\label{eq:ldecay}
\end{eqnarray} 
The effective theories of dark mesons that we
consider below will give specific predictions for these couplings. 
We find two qualitatively distinct possibilities: 
\begin{equation}
\boxed{
\begin{array}{rclcc}
\xi & \sim &  1 \;\; & \mbox{``gaugephilic''} \\
\xi & \ll  &  1 \;\; & \mbox{``gaugephobic''} 
\end{array}}
\end{equation}
Equation~\eqref{eq:ldecay}, which has been argued purely from custodial symmetry and assumptions about the most relevant connections between the dark sector and the SM, is our first main result.
The main purpose of the rest of this paper is to determine how dark pion interactions in different dark sector theories with (or without) custodial $SU(2)_C$ map into Eq.~\eqref{eq:ldecay} and, especially, whether they fall into the gaugephilic or gaugephobic category. 

Before jumping head first into strongly-coupled dark sectors,
the interactions of a custodial $SU(2)$ triplet given in
Eq.~(\ref{eq:ldecay}) are perhaps most familiar from 
two-Higgs doublet models.  We take a brief look 
at this in the next section, leaving a detailed discussion 
to Appendix~\ref{app:2HDM}. 
 
\subsection{Two-Higgs Doublet Models}

As a point of reference, it is helpful to consider the couplings 
of $(H^\pm,A^0)$ in two-Higgs doublet models (2HDMs).  The couplings
to the fermions are model-dependent; for illustration here
let's consider the so-called Type~I 2HDM where the fermions
couple to just one Higgs doublet as that is the 2HDM setup that 
most closely resembles the our dark pion theories.  
In Type~I 2HDM theory, one obtains \cite{Craig:2012pu} 
\begin{eqnarray}
\frac{1}{v_\pi} &=& \frac{1}{v} \cot \beta \nonumber 
\end{eqnarray} 
where we have neglected the CKM mixing for the charged Higgs couplings. 
For the gauge/Higgs sector, 
\begin{eqnarray} 
\frac{\xi}{v_\pi} &=& \frac{1}{v} \cos(\beta - \alpha) \nonumber 
\end{eqnarray} 
Here, $\cot\beta$ is the usual ratio of the expectation values
in two Higgs doublets. In the decoupling limit, the coupling to 
gauge/Higgs boson is well-known to scale as \cite{Gunion:2002zf} 
\begin{eqnarray}
|\cos(\beta - \alpha)| &\sim& \frac{v^2}{m_A^2} \, . 
\end{eqnarray}
Since the coupling of fermions does not have a similar scaling,
we see that 2HDMs are gaugephobic regardless of the Type of 2HDM\@. 

There is an interesting story about utilizing the custodially-symmetric
basis for 2HDMs.  
In the decoupling limit a 2HDM becomes custodially symmetric, 
and the decays of its heavy states ($H^\pm,A^0$) to SM particles 
in this limit are gaugephobic. 
Details are presented 
in Appendix~\ref{app:2HDM}. 

\subsection{Neutral Dark Pion Decay to Diphotons}
 
Finally, it is interesting to discuss the coupling
of $\pi^0$ to $\gamma\gamma$.  
The usual axial anomaly contribution to this decay mode,
$\pi^0 F_{\mu\nu} \widetilde{F}^{\mu\nu}/f$,
whose leading contribution is proportional to 
${\rm Tr} \, Q^2 t^3_a$ [where $t^3_a$ is the generator of the 
axial $U(1)$]
is conspicuously absent from Eq.~\eqref{eq:ldecay}. 
The reason for this is that in a dark sector where the $SU(N)_V$, 
preserved by strong interactions, is an exact symmetry, 
this contribution must vanish. 
For example, in a two-flavor dark sector, invariance under an exact $SU(2)_V$ would enforce the two flavors of dark fermion masses are equal.  Gauging the full $SU(2)_V$
[as in $SU(2)_L$] or just the $t^3$ subgroup [as in $U(1)_Y$] implies the dark fermion electric charges are equal and opposite. In this case, ${\rm Tr} \, Q^2 t^3_a$ vanishes, as do higher order $\pi_D^0 \gamma\gamma$ operators proportional to the differences of dark fermion masses. 

Nevertheless, there is a very small, residual contribution to 
$\pi_D^0 \ra \gamma\gamma$, due to the interactions with the 
SM in Eq.~(\ref{eq:ldecay}).  That is, even though custodial 
$SU(2)$ is preserved by the dark sector interactions with the SM, 
the SM itself violates custodial $SU(2)$.  The dark pion
interactions with the SM fermion axial current generate a
one-loop suppressed $\pi_D^0$-$\gamma$-$\gamma$ coupling
proportional to $m_f/(16 \pi^2 v_\pi)$.
We can calculate the amplitude for the rate by borrowing the standard results 
for $A^0$ decay in two-Higgs doublet models \cite{Djouadi:2005gj}
and suitably substituting couplings:
\begin{eqnarray}
\mathcal{A}(\pi^0_D \ra \gamma \gamma) &=& 
  \sum_f
  \frac{\alpha}{4 \pi} N_c Q_f^2 \left(\frac{m_f}{v_\pi} \right) 
  \sqrt{\tau_f} f(\tau_f)
\end{eqnarray}
where $N_c$ is the number of colors, $Q_f$ is the electric charge, 
$\tau_f = 4 m_f^2/m_\pi^2$, and 
\begin{eqnarray}
f(\tau) &=& \left\{ \begin{array}{ll}
    {\rm arcsin}^2 \frac{1}{\sqrt{\tau}} & \tau \ge 1 \\ 
    -\frac{1}{4} \left[ \log \frac{1+\sqrt{1-\tau}}{1-\sqrt{1-\tau}} - i \pi 
                 \right]^2               & \tau < 1 \, .
                    \end{array} \right. 
\end{eqnarray}
In the limit $\tau \ll 1$,
\begin{eqnarray}
f(\tau) \rightarrow - (1/4) [\log (4/\tau) - i \pi]^2 \, ,
\end{eqnarray}
and thus we see an additional suppression of the 
$\pi_D^0$ decay amplitude of roughly 
$\sqrt{\tau_f} = 2 m_f/m_{\pi_D}$ 
when $m_{\pi_D} \gg m_f$ (neglecting the $\tau$ dependence of the log).  
Hence, while there is $\pi_D^0$ decay to 
$\gamma\gamma$ due to the custodial $SU(2)$ breaking in the SM,  
the decay rate is suppressed by roughly 
$\alpha^2/(16 \pi^2) \times (4 m_f^2/m_{\pi_D}^2)$ 
that is $\simeq 10^{-6} \times m_f^2/m_{\pi_D}^2$ smaller than the 
direct decay to fermions.  This is so small as to be 
phenomenologically irrelevant.

\section{Two-flavor theories}
\label{sec:twoflavor}

The simplest anomaly-free dark sector theories that we consider
have two flavors of dark fermions. We refer to the dark color 
fundamentals as $F_i$ and anti-fundamentals as  $\hat F_i$
transforming under $SU(N_D)$ with flavor index $i = 1,2$. 
The global symmetry of the flavors is 
$SU(2)_{\rm fund} \times SU(2)_{\rm anti}$. Once we include 
interactions between the dark fermions and the Higgs multiplet, 
we will be forced to connect the fermion flavor symmetries to the 
$SU(2)_L \times SU(2)_R$ symmetry of the Higgs potential. 
This connection can be made in a few different ways, two vector-like 
and one chiral. In the vector-like assignments, both $F_i$ and $\hat F_i$ 
must be doublets of the same $SU(2)$ -- either $SU(2)_L$ or $SU(2)_R$, 
while in the chiral assignment, $F$ and $\hat F$ transform under 
different $SU(2)$s.\footnote{Anomaly
cancellation requires $N_D$ to be even for the chiral case.} 
However, in all of these cases, $SU(2)_{\rm fund} \times SU(2)_{\rm anti}$ 
is broken to the diagonal $SU(2)_V$ by strong dynamics, just as in 
two-flavor QCD\@.  Also just like QCD, the dark pions form a triplet 
of the diagonal $SU(2)_V$, which ultimately becomes 
$(\pi^+, \pi^0, \pi^-)$ after electroweak breaking down to just 
$U(1)_{\rm em}$.  This is the custodial $SU(2)$ symmetric triplet 
that we discussed in the previous section.

Prior to electroweak breaking scale, all three pions
$\pi^\pm,\pi^0$ are stable.  Once electroweak symmetry is 
broken, electromagnetic corrections split the multiplets by \cite{Das:1967it} 
\begin{eqnarray}
m_{\pi^\pm}^2 - m_{\pi^0}^2 &=& \frac{(3 \ln 2)}{2 \pi} \, \alpha \, m_\rho^2 
\label{eq:EW_splitting} 
\end{eqnarray} 
where $\alpha$ is the electroweak coupling constant, and $m_\rho$ 
is the mass of the vector resonances of the dark sector. 
This mass splitting allows the weak decay of 
$\pi^\pm \ra \pi^0 \bar{f}' f$. 
Whether this decay is competitive (or not) with direct decays
$\pi \ra {\rm SM}$ will depend on the $\pi$-${\rm SM}$-${\rm SM}$
coupling strength proportional to $1/v_\pi$ in the effective theory.

We now consider each of these theories in turn.   

\subsection{Two-flavor chiral theory}
\label{twoflavorchiral-sec}

\begin{table}[h!]
\begin{center}
\begin{tabular}{c|c}
Field & $(SU(N_D), SU(2)_L, SU(2)_R)$ \\ \hline
$F$ & $(\mathbf{N}, \mathbf{2}, \mathbf{1})$  \\
$\hat F$ & $(\overline{\mathbf{N}}, \mathbf{1}, \mathbf{2})$  \\
\end{tabular}
\end{center}
\caption{Two-flavor fermion content of the chiral theory.}
\label{twoflavorchiral:table}
\end{table}

The two-flavor chiral theory contains the matter content
in Table~\ref{twoflavorchiral:table}.  
$SU(2)_L$ is embedded as $SU(2)_{\rm fund}$ while 
$U(1)_Y$ is the $t^3$ generator of $SU(2)_{\rm anti}$. 
Confinement breaks the global symmetry to $SU(2)_V$, of which only
the gauged $U(1)_{\rm em}$ survives.  

Identifying the flavor symmetries $SU(2)_{fund}, SU(2)_{anti}$ with $SU(2)_L, SU(2)_R$ respectively, we can write a Yukawa interaction between the Higgs bi-doublet and the dark fermions  
\begin{eqnarray}
\mathscr{Y}_{\rm Yuk} &=& y F \mathcal{H} \hat F + h.c. \, ,
\label{eq:chiralYuk}
\end{eqnarray}
Once the Higgs acquires a vev, this will give gives equal
contributions to the masses
of the ``up-type'' and ``down-type'' dark fermions. In the absence of a fundamental Higgs, this theory is minimal technicolor. Including the Higgs (and Yukawa coupling), the two-flavor chiral theory dynamics ``induces'' electroweak symmetry 
breaking even when the Higgs multiplet (mass)$^2$ is positive.
This theory is better known as bosonic technicolor~\cite{Simmons:1988fu,Samuel:1990dq} 
or strongly-coupled induced electroweak symmetry breaking~\cite{Azatov:2011ht, Azatov:2011ps}.

Now that we have established how dark fermions transform under 
$SU(2)_L \times SU(2)_R$ we can consider a more general set of 
interactions that arise with higher dimensional operators. 
These terms can involve more Higgs fields, derivatives, 
SM quarks and/or leptons. Examples at dimension-6 include:
\begin{equation}
c_{6A}\,\frac{(F\, \hat F)(Q_L \hat Q_R)}{\Lambda^2} , \;
c_{6B}\,\frac{(F^{\dag}\bar{\sigma}^{\mu}\, F)
         (\mathcal{H}\, D_{\mu} \mathcal{H})}{\Lambda^2}, \;
c_{6C}\,\frac{(F^{\dag}\bar{\sigma}^{\mu}\,t^a_L\, F)
           (\mathcal{H}\, t^a_L\, D_{\mu} \mathcal{H})}{\Lambda^2}, \cdots
\label{eq:chiralhigherdimoperators}
\end{equation}
where $t^a_L$ are the generators of $SU(2)_L$ that pick out the 
triplet combination of the two doublets.  We will use $t^a_L$ 
and the $SU(2)_R$ counterpart $t^a_R$ throughout this paper.

The translation to the NLSM involves
\begin{eqnarray}
F \hat F \to 4\pi f^3 \Sigma, \quad 
\Sigma = \exp\left[ i \frac{2 \pi^a t^a}{f} \right] \, . 
\end{eqnarray}
The covariant derivative acts on $\Sigma$ identically to 
the Higgs bi-doublet, Eq.~(\ref{eq:higgs2x2covar}),
leading to interactions of the dark pions with the 
electroweak gauge bosons. While there is a systematic way to transmute
interactions between a strong, chiral symmetry breaking sector and 
external fields into interactions involving
pNGBs~\cite{Callan:1969sn,Bando:1987br}, we do not need the 
full machinery since we are
interested in the additional (higher dimensional) terms in the dark sector
chiral lagrangian that, after expanding $\Sigma$,
involve a single power of $\pi^a$.  This criteria selects out operators 
whose dark sector components are i.) Lorentz invariant, as we want operators 
with $\pi_D$, not $\partial_{\mu}\pi_D$, and ii.) that transform 
non-trivially under $SU(2)_C$ -- as discussed in Sec.~\ref{sec:triplets}, 
the dark pion decay terms involve connecting $SU(2)_C$ triplets in the 
strong sector with SM $SU(2)_C$ triplets. In the chiral case, 
these criteria tell us to ignore operators containing
$F^{\dag}\bar{\sigma}^{\mu} F$ (inert under $SU(2)_C$ and not a 
Lorentz invariant) in favor of operators containing $F \hat F$.

Performing the translation to pNGB form and focusing on the most 
relevant interactions between the dark fermions and the Higgs/SM, 
the theory becomes
\begin{eqnarray}
\mathscr{L} &=& \frac{f^2}{4} \mathrm{Tr} \, 
                 (D_\mu \Sigma)^\dagger D^\mu \Sigma \nonumber \\
& &{} + 4 \pi f^3 y \, \mathrm{Tr} \, (\mathcal{H} \Sigma^\dagger  
      + h.c.) \, + \, \text{higher dimensional terms}\, . 
\end{eqnarray}
Here $\Sigma$ contains the triplet of dark pions, and ``higher dimensional'' here refers to operators such as Eq.~\eqref{eq:chiralhigherdimoperators} that are non-renormalizable when written in the UV, in terms of the underlying dark fermions.  In this model the higher dimensional operators are subdominant
(and so we can ignore them for now), but as we will see in later sections, in other models they are vital to connect the dark sector to the SM\@. 

With the pNGB description of the theory in hand, we can now work out how these $\Sigma$ interactions map into interactions among dark pions to SM fields in Eq.~\eqref{eq:ldecay}. 
The term linear in $\Sigma$ expresses the explicit 
chiral symmetry breaking that arises from the 
Yukawa interactions.  
Expanding the linear term out to quadratic order in the 
dark pion fields, 
\begin{eqnarray}
 4 \pi f^3 y \, \mathrm{Tr} \, (\mathcal{H} \Sigma^\dagger + h.c.) 
&\supset& 8 \pi f^2 y \, G^a \pi^a + 4 \pi f y v \, \pi^a \pi^a \, ,
\end{eqnarray}
we see the dark pions acquire masses $m_\pi^2 = 8 \pi f y v$
and mixing between the would-be Goldstones of the
Higgs doublet and the triplet of dark pion fields.
The Goldstone-dark pion mixing is independent of $a$, 
as it should be given the custodially-symmetric origin
of the Yukawa couplings. 

Defining the physical pions and Goldstones as
\begin{eqnarray}
\left( \begin{array}{c}
G^a_{\rm phys} \\
\pi^a_{\rm phys}
\end{array} \right)
&=& V
\left( \begin{array}{c}
G^a \\
\pi^a 
\end{array} \right)
\end{eqnarray}
where the mixing angle is determined by diagonalizing the
mass matrix
\begin{eqnarray}
M_{\rm diag}^2 &=& V M^2 V^T
\end{eqnarray}
with 
\begin{eqnarray}
M^2 &=& \left( \begin{array}{cc} 8 \pi f^3 y/v & 8 \pi f^2 y \\
                               8 \pi f^2 y & 8 \pi f y v \end{array} \right) 
\quad , \quad 
V \; = \; \left( \begin{array}{cc} c_\theta & - s_\theta \\
                                   s_\theta & c_\theta \end{array} \right) 
\end{eqnarray}
and $\theta = {\rm arctan} (f/v)$ is the mixing angle.
The nonzero entry for Goldstone part of the 
mass matrix ($G^a G^a$) arises after minimizing the 
Higgs potential to include the contributions from the dark sector
(see \cite{Chang:2014ida} for details).
Inserting the diagonalized eigenstates back into 
the Lagrangian leads to a shift of the electroweak vev
\begin{equation}
v^2 + f^2 = v_{\rm 246}^2 \simeq (246 \; {\rm GeV})^2 \, .
\label{eq:vevchiral} 
\end{equation}
This leads to well-known corrections to 
Higgs couplings~\cite{Chang:2014ida}.  For our purposes, 
the couplings of the physical pions to $\bar{f} f$, $Z h$ and 
$\bar{f}' f$, $W h$ become
\begin{eqnarray}
(\pi^\pm_{\rm phys} \partial_\mu h - h \partial_\mu \pi^\pm_{\rm phys})  W^{\mu,\mp}: 
& & \frac{M_W}{v} s_\theta \nonumber \\ 
(\pi^0_{\rm phys} \partial_\mu h - h \partial_\mu \pi^0_{\rm phys}) Z^\mu: 
& & \frac{M_Z}{v} s_\theta \nonumber \\
\pi^\pm_{\rm phys} \bar{f}' f: 
& & \sqrt{2} \, \left( \frac{m_{f'}}{v} P_L - \frac{m_f}{v} P_R \right) 
(2 T_3^f) \, s_\theta \nonumber \\ 
\pi^0_{\rm phys} \bar{f} f: 
& & i \, \left( \frac{m_f}{v} \gamma_5 \right) (2 T_3^f) \, s_\theta 
\end{eqnarray} 
where $2 T_3^f = \pm 1$ is the isospin of the fermion.  
The mixing angle is 
\begin{eqnarray}
s_\theta &=& \frac{f}{v_{\rm 246}} \, .
\end{eqnarray}
We see that custodially-symmetric two-flavor chiral theories
have couplings to fermions and gauge bosons that are 
parametrically comparable -- $M_{W,Z}$ versus $m_f$. 
From the couplings we can identify
\begin{eqnarray}
\frac{1}{v_\pi} \simeq \frac{1}{v} \, \times \, 
                       \left( \frac{f}{v_{\rm 246}} \right) \, , 
\qquad \xi = 1 
\end{eqnarray}
so the couplings are ``gaugephilic'' according to Eq.~\eqref{eq:ldecay}. 
While this provides an excellent example of ``gaugephilic'' 
dark pion interactions, there is no way to formally 
separate the Goldstone/pion mixing from the dark pion mass 
itself -- both are proportional to the Yukawa coupling $y$. 
Consequently, there is no limit where the mixing between
the Goldstone and the dark pion can be taken small while simultaneously
holding the dark pion mass fixed.

We should emphasize that in the two-flavor chiral model we arrive 
at Eq.~\eqref{eq:ldecay} through the mixing of the dark pions 
with the triplet of Goldstone bosons. This mixing was possible only because 
of the Yukawa term, which is the only allowed renormalizable coupling. Had we included the higher dimensional terms in the chiral lagrangian, 
we would find that they can still be parameterized by the
effective lagrangian Eq.~(\ref{eq:ldecay}).
In two-flavor vector-like models, which we explore next, the 
dark pion--Goldstone mixing is not present, however we will still recover 
Eq.~\eqref{eq:ldecay}.

Finally, the absence of $\pi^0$-$\gamma$-$\gamma$ coupling
critically relied on the renormalizable coupling between the 
dark sector and the SM, Eq.~(\ref{eq:chiralYuk}), being 
custodially symmetric.  If there had been an explicit 
custodial violation of the dark sector with Higgs multiplet, e.g.,
$y^{\slashed{C}} F \mathcal{H} t^3_R \hat F$, the pions would acquire
different masses as well as different mixings with the
Goldstones.  This would re-introduce $\pi^0 \ra \gamma\gamma$
and a more detailed calculation would be needed to determine
the branching fractions of $\pi^0$. 

\subsection{Two-flavor vector-like theories}
\label{sec:twoflavorcustodial}

Vector-like confinement~\cite{Kilic:2009mi} popularized the possibility
that a new strong sector contains fermions in vector-like
representations so that contributions to electroweak precision
corrections are negligible, and (bare) vector-like masses for 
the dark fermions are allowed.  
There are two versions of two-flavor vector-like theories, 
shown in Table~\ref{twoflavorvectorlike:table},
depending on whether the dark fermions transform under
just $SU(2)_L$ or just $SU(2)_R$.  We will refer to these as the 
``$SU(2)_L$ model'' and ``$SU(2)_R$ model'', respectively.

\begin{table}[h!]
\begin{center}
\begin{tabular}{c|c}
\multicolumn{2}{c}{$SU(2)_L$ model} \\
\\
Field & $(SU(N_D), SU(2)_L, SU(2)_R)$ \\ \hline
$F$ & $(\mathbf{N}, \mathbf{2}, \mathbf{1})$  \\
$\hat F$ & $(\overline{\mathbf{N}}, \mathbf{2}, \mathbf{1})$  \\
\end{tabular}
\hspace*{1cm}
\begin{tabular}{c|c}
\multicolumn{2}{c}{$SU(2)_R$ model} \\
\\
Field & $(SU(N_D), SU(2)_L, SU(2)_R)$ \\ \hline
$F$ & $(\mathbf{N}, \mathbf{1}, \mathbf{2})$  \\
$\hat F$ & $(\overline{\mathbf{N}}, \mathbf{1}, \mathbf{2})$  \\
\end{tabular}
\end{center}
\caption{Two-flavor fermion content of $SU(2)_L$ and $SU(2)_R$ 
vector-like theories.}
\label{twoflavorvectorlike:table}
\end{table}

Vector-like theories permit dark fermion masses,
\begin{eqnarray}
\mathscr{L}_{\rm mass} &=& M F \hat F + h.c. \, .
\label{eq:masstwoflavorvectorlike}
\end{eqnarray}
The global $SU(2)_{\rm fund} \times SU(2)_{\rm anti}$ symmetries 
are broken to $SU(2)_V$ that is identified either with 
the fully gauged $SU(2)_L$ or $SU(2)_R$ (with, as usual, 
just $U(1)_Y$ gauged). 

Now we begin to add interactions between the dark fermions and the 
SM fields, working in a $SU(2)_L \times SU(2)_R$ invariant manner. 
Unlike the two-flavor chiral model, in the vector-like models 
we cannot write a renormalizable interaction between 
$F, \hat F$ and $\mathcal{H}$.  
To write down interactions between the Higgs and the dark fermions, 
we need to consider higher dimensional operators.
Both the $SU(2)_L$ and $SU(2)_R$ vector-like models
theories allow the ``singlet'' contribution at dimension-5
\begin{eqnarray}
c_{5M} \frac{(F \hat F) \mathrm{Tr} \, \mathcal{H}^\dagger \mathcal{H}}{\Lambda}
+ h.c. \, 
\label{eq:dim5singlet}
\end{eqnarray}
The $(F \hat F)$ part and the 
$\mathrm{Tr} \, \mathcal{H}^\dagger \mathcal{H}$
are singlets that are separately invariant under their 
respective global symmetries.  After electroweak symmetry breaking, 
this operator leads to a $\sim v^2/\Lambda$ contribution to the 
dark fermion masses but does not influence their decays.  
This is because the first non-zero interactions arising from 
expanding out Eq.~\eqref{eq:dim5singlet} must contain a singlet,
i.e., at least two dark pions.

Hence, to find an operator contributing to dark pion decay we need 
to go beyond dimension-5.
We seek a non-singlet contraction of $F$ and $\hat F$.
In the case of the $SU(2)_L$ model, this is 
$F t^a_L \hat{F}$.  In the case of the $SU(2)_R$ model,
invariance under the full $SU(2)_R$ allows just 
$F t^a_R \hat{F}$.  Of course given that just $U(1)_Y$ is gauged,
the term $F t^3_R \hat{F}$ is gauge-invariant but not 
$SU(2)_R$ invariant.  If we insist that the dark sector
preserves custodial $SU(2)$, this combination is forbidden.

In the Standard Model, there are no dimension-3 operators of 
the form $Q t^a_L Q'$ since, of course, the SM fermions 
transform under a chiral representation of the electroweak group.
By dimension-4 we can write, e.g., $Q_L t^a_L \mathcal{H} \hat Q_R$, 
which can be combined with the $F t^a_{L,R} \hat{F}$ from the dark sector 
to obtain dimension-7 operators including:
\begin{eqnarray}
SU(2)_L \; {\rm model:} \qquad  \mathscr{L} &=& 
\mathcal{Y}^C_{ij}
\frac{(F t^a_L \hat F) ({Q_L}_i t^a_L \mathcal{H} \frac{\mathds{1}_R}{2} 
\hat {Q_R}_j)}{\Lambda^3}
+ \mathcal{Y}^{\slashed{C}}_{ij} 
\frac{(F t^a_L \hat F) ({Q_L}_i t^a_L \mathcal{H} t^3_R 
\hat {Q_R}_j)}{\Lambda^3}
\, , \nonumber \\
SU(2)_R \; {\rm model:} \qquad  \mathscr{L} &=& 
\mathcal{Y}^C_{ij}
\frac{(F t^a_R \hat F) 
({Q_L}_i \mathcal{H} t^a_R \hat {Q_R}_j)}{\Lambda^3}
+ \mathcal{Y}^{\slashed{C}}_{ij} 
\frac{(F t^a_R \hat F) ({Q_L}_i \mathcal{H} t^a_R t^3_R 
      \hat {Q_R}_j)}{\Lambda^3} 
\label{eq:twoflavorfermiondim7}
\end{eqnarray}
As we discussed in Sec.~\ref{sec:custodial},
we have included the SM Yukawa couplings
as coefficients to these operators in order to 
maintain minimal flavor violation.

Focusing on just one generation of SM fermions, 
these dimension-7 operators become 
\begin{eqnarray}
SU(2)_L \; {\rm model:} \qquad \mathscr{L} &=& c_{7f} \frac{(F t^a_L \hat F) 
(Q_L t^a_L \mathcal{H} Y_{ud} \hat Q_R)}{\Lambda^3} \, , \nonumber \\
SU(2)_R \; {\rm model:} \qquad \mathscr{L} &=& c_{7f} \frac{(F t^a_R \hat F) 
(Q_L \mathcal{H} t^a_R Y_{ud} \hat Q_R)}{\Lambda^3} \, . 
\end{eqnarray}
where $Y_{ud}$ is a $2\times 2$ matrix in $SU(2)_R$ space
with the form $Y_{ud} = (y_u + y_d) \mathds{1}_R/2 + (y_u - y_d) t^3_R$.
After electroweak symmetry breaking, this operator mixes 
$(F t^a_{L,R} \hat F)$ with a triplet combination of 
SM fermions  
$(y_d u_L d_R^c, \; y_u u_L u_L^c - y_d d_L d_L^c, \; y_u d_L u_L^c)$.
Passing to the non-linear sigma model formalism, the dimension-7 operator 
becomes
\begin{eqnarray}
SU(2)_L \; {\rm model:} \qquad
\mathscr{L} &=& c_{7f} \frac{4 \pi f^3}{\Lambda^3} 
\left( \mathrm{Tr} \Sigma_L t^a_L \right)  
Q_L t^a_L \mathcal{H} Y_{ud} \hat Q_R \,  \nonumber \\
SU(2)_R \; {\rm model:} \qquad
\mathscr{L} &=& c_{7f} \frac{4 \pi f^3}{\Lambda^3} 
\left( \mathrm{Tr} \Sigma_R t^a_R \right)  
Q_L \mathcal{H} t^a_R Y_{ud} \hat Q_R \, , 
\end{eqnarray}
where $\Sigma_{L,R}$ is in terms of the $SU(2)_{L,R}$ generators,
$\Sigma_{L,R} = \exp[ i 2 \pi^a t^a_{L,R}/f ]$.  
Notice that $\Sigma_L$ transforms as an adjoint
under the $SU(2)_V$ [that is fully gauged as $SU(2)_L$], 
hence the combination $\mathrm{Tr} \Sigma_L t^a_L$ 
expands to $\pi^a/f$ to leading order in $\pi^a$.
Using this expansion, we obtain the interactions: 
\begin{eqnarray}
\pi^\pm_{\rm phys} \bar{f}' f: 
& & \sqrt{2} \, (m_{f'} P_L - m_f P_R) (2 T_3^f) 
\, \times \, (c_{7f} \frac{\sqrt{2} \pi f^2}{\Lambda^3}) \nonumber \\ 
\pi^0_{\rm phys} \bar{f} f: 
& & i \, (m_f \gamma_5) (2 T_3^f) 
\, \times \, (c_{7f} \frac{\sqrt{2} \pi f^2}{\Lambda^3}) 
\end{eqnarray} 
From this we can identify 
\begin{equation} 
\frac{1}{v_\pi} = c_{7f} \frac{\sqrt{2} \pi f^2}{\Lambda^3} 
\end{equation}
Notice that the interactions are otherwise identical regardless
of whether the underlying theory is $SU(2)_L$ or $SU(2)_R$. 

If we extend the effective theory to even higher dimension operators, 
we encounter operators involving the triplet combination
$F t^a_{L,R} \hat F$ with the Higgs multiplet.  
The lowest dimension operator involving
$F t^a_{L,R} \hat F$ and custodially symmetric contractions of
of powers of $\mathcal{H}$ occurs at dimension-9:
\begin{eqnarray}
SU(2)_L \; {\rm model:} \qquad 
\mathscr{L} &=& 
c_{9C}  \epsilon_{abc} \delta_{de} \frac{(F t^a_L \hat F) 
        \mathrm{Tr} \left[
        (D_\mu \mathcal{H})^\dagger t^b_L (D^\mu \mathcal{H}) t^d_R 
        \mathcal{H}^\dagger t^c_L \mathcal{H} t^e_R \right]}{\Lambda^5} \; 
        \nonumber \\
SU(2)_R \; {\rm model:} \qquad 
\mathscr{L} &=& 
c_{9C}  \epsilon_{abc} \delta_{de} \frac{(F t^a_R \hat F) 
        \mathrm{Tr} \left[
        (D_\mu \mathcal{H})^\dagger t^d_L (D^\mu \mathcal{H}) t^b_R 
        \mathcal{H}^\dagger t^e_L \mathcal{H} t^c_R \right]}{\Lambda^5} 
\label{eq:dim7custodiallypreserving}
\end{eqnarray}
Passing to the low energy effective theory, the non-linear sigma model
acquires the same kinetic and mass terms as in
Eq.~(\ref{eq:nlsmcustodialviolating}) with an interaction term
\begin{eqnarray}
SU(2)_L \; {\rm model:} \quad 
\mathscr{L} &=&  
c_{9C} \frac{4 \pi f^3}{\Lambda^5} \epsilon_{abc} \delta_{de} 
\mathrm{Tr} \left[ \Sigma_L t^a_L \right] \mathrm{Tr} \left[  
(D_\mu \mathcal{H})^\dagger t^b_L (D^\mu \mathcal{H}) t^d_R
\mathcal{H}^\dagger t^c_L \mathcal{H} t^e_R \right] 
\nonumber \\
SU(2)_R \; {\rm model:} \quad 
\mathscr{L} &=&  
c_{9C} \frac{4 \pi f^3}{\Lambda^5} \epsilon_{abc} \delta_{de} 
\mathrm{Tr} \left[ \Sigma_R t^a_R \right] \mathrm{Tr} \left[ 
(D_\mu \mathcal{H})^\dagger t^d_L (D^\mu \mathcal{H}) t^b_R
\mathcal{H}^\dagger t^e_L \mathcal{H} t^c_R \right] 
\end{eqnarray} 
where $\Sigma_{L,R}$ is as before.
Expanding the interaction in unitary gauge to leading order 
in $\pi^a$ we obtain: 
\begin{equation}
\mathscr{L} \; = \; 
c_{9C} \frac{\pi f^2}{16 \Lambda^5} (v+h)^3 \left[
g W^\mp_\mu \left( \pi^\pm \partial^\mu h - h \partial^\mu \pi^\pm \right) 
+ \sqrt{g^2 + {g'}^2} Z_\mu 
\left( \pi^0 \partial^\mu h - h \partial^\mu \pi^0 \right) \right] 
\label{eq:twoflavorcustodialcouplings}
\end{equation}
and thus the couplings are 
\begin{eqnarray}
(\pi^\pm_{\rm phys} \partial_\mu h - h \partial_\mu \pi^\pm_{\rm phys}) 
W^{\mu,\mp}: 
& & M_W \, \times \, \left( c_{9C} \frac{\pi f^2}{\Lambda^3} \right) 
\, \times \, \left( \frac{v^2}{8 \Lambda^2} \right) \nonumber \\ 
(\pi^0_{\rm phys} \partial_\mu h - h \partial_\mu \pi^0_{\rm phys}) Z^\mu: 
& & M_Z \, \times \, \left( c_{9C} \frac{\pi f^2}{\Lambda^3} \right) 
\, \times \, \left( \frac{v^2}{8 \Lambda^2} \right) 
\end{eqnarray}
Compare to the fermion couplings, we then obtain 
\begin{eqnarray}
\frac{1}{v_\pi} = c_{7f} \frac{\sqrt{2} \pi f^2}{\Lambda^3} \; , \qquad 
\xi = \left( \frac{c_{9C}}{c_{7f}} \right) \, \times \, 
      \left( \frac{v^2}{8 \sqrt{2} \Lambda^2} \right) \, .
\end{eqnarray}
The single dark pion interactions with the Standard Model 
can be precisely characterized by the effective lagrangian 
Eq.~(\ref{eq:ldecay}).  Unlike the two-flavor chiral model, 
in the vector-like models there is no Goldstone/dark pion mixing 
connecting the dark sector with the Standard Model.  
Instead, this is fully characterized by the higher dimensional 
interactions that, by assumption, preserve custodial $SU(2)$.  

Notice also that the coefficient of the 
$\pi \mbox{-} V \mbox{-} h$
interaction is \emph{suppressed} relative to the 
$\pi \mbox{-} f \mbox{-} f$ interaction
by an amount $\xi \propto v^2 / \Lambda^2$. In this particular model,
the suppression arises because custodial symmetry demanded that
operators involving the Higgs multiplets appear at a dimension that 
is \emph{two powers} higher than that for SM fermions.
Thus, dark pions preferentially interact with (and ultimately decay
primarily to) SM fermions -- these theories are gaugephobic -- 
in two-flavor, vector-like, custodially-preserving dark sector 
theories. \\

\section{Four-flavor theories}
\label{sec:fourflavor}

The main disadvantage to limiting ourselves to two flavors
of fermions is that we are forced to choose between 
either having chiral masses or vector-like masses for 
fermions at the renormalizable level.  With four flavors,
we can engineer the electroweak quantum numbers to
permit both vector-like and chiral masses, governed by
Lagrangian parameters that are fully adjustable.

Large chiral masses with small vector-like masses 
will tend to cause the dark sector to substantially
break electroweak symmetry (and violate bounds from 
the $S$ parameter as well as Higgs coupling measurements).  
Therefore, we focus on the opposite case -- the parameter space where 
the dark sector fermion masses are \emph{mostly} vector-like with 
small chiral masses where $y v/M \ll 1$.  
In this way, these theories are automatically safe
from electroweak precision constraints and Higgs coupling
measurements. 
Yet, the presence of both vector-like and small chiral masses
in general means that the dark sector flavor symmetries
are broken to $SU(2)_L \times SU(2)_R \times U(1)_{\rm dark \; baryon}$. 
The existence of baryons stabilized by the accidental
$U(1)_{\rm dark \; baryon}$ was
exploited by the Stealth Dark Matter model \cite{Appelquist:2015yfa}. 
In that theory, with $N (\ge 4, \mbox{even})$, 
the lightest baryon was shown to be a viable dark 
matter candidate.  In this paper, we focus solely on the 
mesons of the dark sector that was of only peripheral interest
in the dark matter papers. 

\begin{table}
\begin{center}
\begin{tabular}{c|c}
Field & $(SU(N_D), SU(2)_L, SU(2)_R)$ \\ \hline
$F_L$ & $(\mathbf{N}, \mathbf{2}, \mathbf{1})$  \\
$\hat F_L$ & $(\overline{\mathbf{N}}, \mathbf{2}, \mathbf{1})$  \\
$F_R$ & $(\mathbf{N}, \mathbf{1}, \mathbf{2})$  \\
$\hat F_R$ & $(\overline{\mathbf{N}}, \mathbf{1}, \mathbf{2})$  \\
\end{tabular}
\end{center}
\caption{Four-flavor, custodially-symmetric dark sector fermion content.}
\label{table:fourflavor}
\end{table}

The field content of our prototype four-flavor,
custodially-symmetric theory is given in
Table~\ref{table:fourflavor}.  
At dimension-3, the vector-like masses for the dark fermions are
\begin{eqnarray} 
\mathscr{L} = M_{12} F_L \hat{F}_L + M_{34} F_R \hat{F}_R + h.c. \, .
\end{eqnarray}
At dimension-4, the chiral masses for the dark fermions are
\begin{eqnarray}
\mathscr{L} = y_{14} F_L \mathcal{H} \hat{F}_R 
              + y_{23} \hat{F}_L \mathcal{H} F_R 
              + h.c. \, .
\end{eqnarray}
With fully general $M_{12}$, $M_{34}$, $y_{14}$, $y_{23}$, 
and the gauging of $SU(2)_L \times U(1)_Y$, 
we see that vector-like and chiral masses arise at the
renormalizable level, unlike the case of the 
two-flavor theories.\footnote{We have switched notation 
$(F_1,F_2,F_3,F_4) \ra (F_L,\hat{F}_L,F_R,\hat{F}_R)$
but retained the same mass and Yukawa coupling parameter
names as Ref.~\cite{Appelquist:2015yfa}.}
We could also include higher dimensional operators 
that we considered earlier in Sec.~\ref{sec:twoflavor}.
But like the two-flavor chiral theory, 
we anticipate the renormalizable interactions
with the SM Higgs sector above will dominate over the 
higher dimensional ones, and so we won't consider them 
further in this section.

After electroweak symmetry breaking, the mass matrix
for the dark fermions can be written in a fully 
Higgs field-dependent way as
\begin{eqnarray}
\mathscr{L}_{\rm mass} = - 
( \;\; F_L^u \, -i F_L^d \;\; F_R^u \, -i F_R^d \, ) \mathcal{M}
\left( \begin{array}{c} 
       \hat{F}_L^d \\
       -i \hat{F}_L^u \\
       \hat{F}_R^d \\
       -i \hat{F}_R^u
       \end{array} \right) + h.c. \, ,
\label{eq:bigmasses}
\end{eqnarray}
where
\begin{eqnarray}
\mathcal{M} = 
\left( \begin{array}{cccc}
       M_{12} & 0      & \frac{y_{23} (-i G^0 + h + v)}{\sqrt{2}} & -i y_{23} G^+ \\
       0      & M_{11} & -i y_{23} G^- & \frac{y_{23} (i G^0 + h + v)}{\sqrt{2}} \\
       \frac{y_{14} (i G^0 + h + v)}{\sqrt{2}} & i y_{14} G^+  & M_{34} & 0 \\
       i y_{14} G^- & \frac{y_{14} (-i G^0 + h + v)}{\sqrt{2}} & 0 & M_{34} 
       \end{array} \right)
\label{eq:bigmassmatrix}
\end{eqnarray}

The field-independent mass terms break up into two $2 \times 2$
mass matrices -- one for the $Q = +1/2$ fermions and one for the 
$Q=-1/2$ fermions that are identical due to custodial symmetry.
It is very convenient to rewrite 
$y_{14} = y (1 + \eps)$ and $y_{23} = y (1 - \eps)$
since, as we will see, contributions to electroweak precision 
observables is proportional to $(\eps y)^2$.  Using this
parameterization, the $2 \times 2$ mass matrices are
\begin{eqnarray}
M_u \; = \; M_d &=&
\left( \begin{array}{cc}
       M_{12} & y (1-\eps) v/\sqrt{2} \\
       y (1+\eps) v/\sqrt{2} & M_{34} 
       \end{array} \right) \, .
\end{eqnarray}
The mass matrix can be diagonalized by a biunitary
transformation involving
\begin{eqnarray}
\tan 2\theta_1 &=& - \frac{\sqrt{2} y v \left( \Delta \eps - M \right)}{
                           2 \Delta M + \eps y^2 v^2} \nonumber \\
\tan 2\theta_2 &=& \frac{\sqrt{2} y v \left( \Delta \eps + M \right)}{
                           2 \Delta M - \eps y^2 v^2}
\label{eq:mixingangles}
\end{eqnarray}
where $M \equiv (M_{12} + M_{34})/2$, $\Delta \equiv (M_{34} - M_{12})/2$, 
and $\theta_1$ ($\theta_2$) diagonalizes $M_u M_u^T$ ($M_u^T M_u$).
The diagonalized fermion masses are
\begin{eqnarray}
m_{1,2} &=& M \mp \sqrt{\Delta^2 + \frac{y^2 (1-\eps^2) v^2}{2}} \, .
\label{eq:darkfermionmasseigenstates}
\end{eqnarray}
We can use these results to rotate Eq.~(\ref{eq:bigmasses})
into the mass basis.  The field-independent parts of the
mass matrix are, of course, fully diagonalized.  But the 
field-dependent ones are not.  We need the field-dependence 
to determine the dark pion / Goldstone mixing.

Passing to the non-linear sigma model, we use
\begin{equation}
\Sigma = \exp \left[ \frac{2\,i \pi^a\, t^a_{15}}{f} \right] \, ,
\end{equation}
where the $\pi^a$ are in the adjoint representation of $SU(4)_V$. Decomposing $\pi^a$ into multiplets of $SU(2)_L \times SU(2)_R$, we have 
\begin{eqnarray}
\mathbf{15} &\to& (\mathbf{3},\mathbf{1}) \oplus 
                  (\mathbf{2},\mathbf{2})_a \oplus 
                  (\mathbf{2},\mathbf{2})_b \oplus 
                  (\mathbf{1},\mathbf{3}) \oplus 
                  (\mathbf{1},\mathbf{1}) \, ,
\end{eqnarray}
where $a$ and $b$ are two separate bi-doublets. 
After rotating into the mass eigenstates of the dark mesons, we have 
\begin{eqnarray}
\Sigma &=& \exp \left[ \frac{i}{f} \left( \begin{array}{cccc} 
\pi_1^0 + \frac{\eta}{\sqrt{2}} & \sqrt{2} \pi_1^+ 
  & K_A^0 & - \sqrt{2} K_B^+ \\
\sqrt{2} \pi_1^- & - \pi_1^0 + \frac{\eta}{\sqrt{2}} 
  & - \sqrt{2} K_A^- & K_B^0 \\
\bar{K}_A^0 & - \sqrt{2} K_A^+ 
  & \pi_2^0 - \frac{\eta}{\sqrt{2}} & \sqrt{2} \pi_2^+ \\ 
- \sqrt{2} K_B^- & \bar{K}_B^0 
  & \sqrt{2} \pi_2^- & - \pi_2^0 - \frac{\eta}{\sqrt{2}} 
\end{array} \right) \right] 
\end{eqnarray}
where we use $\pi_{1,2}$ to denote the dark pions transforming as 
$(\mathbf{3}, \mathbf{1})$ and $(\mathbf{1}, \mathbf{3})$, 
$K_{A,B}$ to denote the ``dark kaons'' that are in 
$(\mathbf{2}, \mathbf{2})$ representations, and
$\eta$ to denote the ``dark eta'' singlet. 

The lowest dimension terms in the NLSM lagrangian are:
\begin{eqnarray}
\mathscr{L}_{\chi} &=& 
  \frac{f^2}{4} \, \mathrm{Tr} (D_{\mu}\Sigma (D^{\mu}\Sigma)^{\dag}) 
+ 4 \pi c_D f^3 \, 
  \mathrm{Tr} \left( L \mathcal{M} R^\dagger \Sigma^\dagger + h.c. \right) \, , 
\label{eq:nlsmfourflavor}
\end{eqnarray}
where $c_D$ is an $\mathcal{O}(1)$ coefficient from the strong dynamics.   As we discussed in Sec.~\ref{twoflavorchiral-sec}, 
these terms are sufficient to 
capture the leading interactions of the dark pions, 
and in particular, will allow us to characterize the single
dark pion interactions with the SM that lead to dark pion decay.
The mixing matrices are formed from the angles Eq.~(\ref{eq:mixingangles})
\begin{eqnarray}
\mathcal{L}
 &=& \left( \begin{array}{cccc}
             \cos\theta_1 &            0 & -\sin\theta_1 & 0 \\
                        0 & \cos\theta_1 &             0 & -\sin\theta_1 \\
             \sin\theta_1 &            0 & \cos\theta_1  & 0 \\
                        0 & \sin\theta_1 &            0 & \cos\theta_1 
             \end{array} \right) \\
\mathcal{R}
  &=& \left( \begin{array}{cccc}
             \cos\theta_2 &            0 & -\sin\theta_2 & 0 \\
                        0 & \cos\theta_2 &             0 & -\sin\theta_2 \\
             \sin\theta_2 &            0 & \cos\theta_2  & 0 \\
                        0 & \sin\theta_2 &            0 & \cos\theta_2
             \end{array} \right) \, . 
\end{eqnarray}
In the field-independent limit,  
\begin{eqnarray}
\mathcal{L} \mathcal{M} \mathcal{R}^\dagger &=& \left( \begin{array}{cccc}
                                   m_1 &   0 &   0 &   0 \\
                                     0 & m_1 &   0 &   0 \\
                                     0 &   0 & m_2 &   0 \\
                                     0 &   0 &   0 & m_2 \\
                                   \end{array} \right) \, , 
\end{eqnarray}
and so the dark pion masses are
\begin{eqnarray}
m_{\pi_1} &=& 4 \pi c_D f (2 m_1)     \label{eq:mpi1} \\ 
m_{K}     &=& 4 \pi c_D f (m_1 + m_2) \label{eq:mK} \\ 
m_{\pi_2} &=& 4 \pi c_D f (2 m_2) \, . \label{eq:mpi2} 
\end{eqnarray}

Finally, the covariant derivative for $\Sigma$ involves the weak currents
\begin{eqnarray}
D_{\mu}\Sigma &=& 
  \partial_{\mu} \Sigma 
  - i\, g W_\mu^{\alpha} \left( j_{\alpha}^V + j_\alpha^A \right) \Sigma 
  - i\, g' B_{\mu} \left( j_Y^V + j_Y^A \right) \Sigma
\end{eqnarray}
where it is convenient to express the vector and axial currents 
explicitly
\begin{eqnarray}
j_{\alpha}^{V,A} &=& \mathcal{L}^\dagger t^\alpha \mathcal{L} 
                     \pm \mathcal{R}^\dagger t^\alpha \mathcal{R} 
                     \quad (\alpha = 1 \ldots 3) \\
j_Y^{V,A} &=& \mathcal{L}^\dagger t^{15} \mathcal{L} 
               \pm \mathcal{R}^\dagger t^{15} \mathcal{R} \, .
\end{eqnarray}
Expanding the covariant derivatives to extract only
the non-derivative contributions -- the mass terms for
$W^\mu$ and $Z^\mu$ -- we find the contributions
of the dark sector to electroweak symmetry breaking
for two flavors:
\begin{eqnarray}
v_{246}^2 &=& v^2 \left( 1 + \frac{\eps^2 y^2 f^2}{M^2} + \ldots \right) \, .
\label{eq:vevstealth}
\end{eqnarray}
Here we have written the leading result in a small 
$\eps$ expansion.  Obviously the correction from
the dark sector, $\eps^2 y^2 f^2/M^2$, should be small to avoid 
constraints from the electroweak precision observables 
as well as Higgs coupling measurements.
In particular, the dark sector's contribution to the $S$ parameter 
can be estimated \cite{Appelquist:2015yfa} utilizing QCD and large $N_D$,
\begin{eqnarray}
S &\sim& \frac{1}{6\pi} N_F N_D \left( \frac{\epsilon y f}{M} \right)^2 
  \simeq 0.1 \frac{N_F}{4} \frac{N_D}{4} 
           \left( \frac{\epsilon y}{0.3} \right)^2 \left( \frac{f}{M} \right)^2
\end{eqnarray}
Since $M < 4 \pi f$ for the NLSM effective theory to be valid,
in general we need $|\eps y|$ small to ensure the dark sector
condensate is aligned nearly (but not completely) in an
electroweak preserving direction.

While Eq.~\eqref{eq:vevstealth} is reminiscent of  Eq.~\eqref{eq:vevchiral} in the two-flavor chiral case, there are some crucial differences. In  Eq.~\eqref{eq:vevchiral}, we could not take $f$ -- the EWSB contribution from the strong sector -- to be arbitrarily small without making the dark pions dangerously light. As a result, there is a minimum $f$ that we can take, and therefore a minimum deviation in Higgs coupling and precision electroweak observables, see Ref.~\cite{Chang:2014ida}. In the four flavor case, we have more freedom. The fact that the fermions are vector-like means we can take $f$ (more correctly $yf$) as small as we like without worrying about $m_{\pi_D}$. This allows us to explore a parameter space where the renormalizable coupling between the Higgs and the dark sector has negligible role on EWSB yet still acts as a portal for the dark pions to decay through.

\subsection{Mixing with the Higgs and Goldstones}

We have chosen a basis for our dark pions such that
they do not acquire an expectation value.
This is evident by expanding the linear term, 
Eq.~(\ref{eq:nlsmfourflavor}), where one finds no terms 
linear in the dark pion fields, i.e., contributions of the form 
$\mathscr{L} \subset ({\rm constant}) \pi$ are absent.

There are, however, dark pion mixing terms with 
both the Higgs field $h$ and (prior to gauge-fixing) 
the Higgs Goldstone fields $G^\pm,G^0$.  
Disentangling the mixing among the Higgs and dark pion
fields is somewhat involved, and in full generality would
need to be done numerically.  In the following, we have calculated
the mixing to leading order in $\eps y$, where we can obtain
analytic expressions.  Since we know $\eps y$ must in general
be small to ensure electroweak symmetry breaking occurs mostly
from the fundamental Higgs field, this is a good choice of an 
expansion parameter.

One unique combination of the dark pion fields mixes
with the Higgs boson $h$,
\begin{eqnarray}
4 \pi c_D f^3 \, 
  \mathrm{Tr} \left( \mathcal{L} \mathcal{M} \mathcal{R}^\dagger \Sigma^\dagger + h.c. \right) 
  & \subset & 4 \sqrt{2} \pi c_D \eps y f^2 h \, \mathrm{Im}( K_A^0 + K_B^0 )
  \, .
\end{eqnarray}
This will turn out to be \emph{critical} to understand the 
effective couplings of the lightest dark pions to the
SM gauge sector.  

The dark pions also mix with the Higgs Goldstones, 
\begin{eqnarray}
\lefteqn{4 \pi c_D f^3 \, 
  \mathrm{Tr} \left( \mathcal{L} \mathcal{M} \mathcal{R}^\dagger \Sigma^\dagger + h.c. \right)
  \; \subset} 
  & & \nonumber \\
& & \frac{8 \pi c_D f^2 \eps y}{M} 
    \bigg[ \Big(  
    G^- \left( s_m (2 m_1 \pi_1^+ - 2 m_2 \pi_2^+) 
               + c_m (m_1 + m_2) (K_A^+ - K_B^+) \right) + h.c. 
    \Big) \nonumber \\
& & \qquad\qquad\quad 
    + G^0 \left( s_m (2 m_1 \pi_1^0 - 2 m_2 \pi_2^0) 
                 + c_m (m_1 + m_2) \mathrm{Re}(K_A^0 - K_B^0) 
    \right) \bigg] \, , 
\end{eqnarray}
where 
\begin{eqnarray}
s_m &\equiv& \sin \theta_m 
             \equiv \frac{\sqrt{2} y v}{\sqrt{2 (y v)^2 + 4 \Delta^2}} \\
c_m &\equiv& \cos \theta_m 
             \equiv \frac{2 \Delta}{\sqrt{2 (y v)^2 + 4 \Delta^2}} \, ,
\end{eqnarray}
are mixing angles among combinations of the dark pions.
The dark pion / Goldstone mass mixing can be perturbatively
diagonalized to leading order in $\eps y$, 
\begin{eqnarray}
G_{\rm phys}^{\pm,0} &=& G^{\pm,0} 
  + \frac{\eps y f}{M} \left( s_m (\pi_1^{\pm,0} - \pi_2^{\pm,0}) 
                              + c_m \, \mathrm{Re}(K_A - K_B) \right) \\
\pi_{1,{\rm phys}}^{\pm,0} &=& \pi_1^{\pm,0} 
                               + \frac{\eps y f}{M} s_m G^{\pm,0} \\
\pi_{2,{\rm phys}}^{\pm,0} &=& \pi_2^{\pm,0} 
                               - \frac{\eps y f}{M} s_m G^{\pm,0} \\
\frac{\mathrm{Re}\left( K_{A,{\rm phys}}^{\pm,0} 
                        - K_{B,{\rm phys}}^{\pm,0} \right)}{\sqrt{2}}
  &=& \frac{\mathrm{Re}\left( K_A^{\pm,0} - K_B^{\pm,0} \right)}{\sqrt{2}} 
      + \frac{\eps y f}{M} c_m G^{\pm,0} \\
\frac{\mathrm{Re}\left( K_{A,{\rm phys}}^{\pm,0} 
                        + K_{B,{\rm phys}}^{\pm,0} \right)}{\sqrt{2}} 
  &=& \frac{\mathrm{Re}\left( K_{A}^{\pm,0} + K_{B}^{\pm,0} \right)}{\sqrt{2}}
  \, .
\end{eqnarray}
In addition, diagonalizing the Higgs boson / dark pion mixing 
one obtains
\begin{eqnarray}
h_{\rm phys} &=& h - \frac{\eps y f m_K}{M} 
                     \frac{\mathrm{Im}( K_A^0 + K_B^0 )}{m_h^2 - m_K^2} \\
\mathrm{Im} \left( K_{A,{\rm phys}}^0 + K_{B,{\rm phys}}^0 \right)  
  &=& \mathrm{Im} \left( K_A^0 + K_B^0 \right) 
       + \frac{\eps y f m_K}{M} \frac{h}{m_h^2 - m_K^2} 
\end{eqnarray}
where $m_K$ is given by Eq.~(\ref{eq:mK}).

\subsection{Dark Pion Couplings to the SM}

We now calculate the couplings of dark pions to the
gauge sector of the Standard Model.  These couplings 
arise when the interaction eigenstates ($G$, $\pi$, $K$) 
are rotated into the physical states ($G_{\rm phys}$, 
$\pi_{\rm phys}$, $K_{\rm phys}$).  Gauge-fixing
in unitary gauge removes all terms involving $G_{\rm phys}$,
leaving just the interactions with the ``physical'' 
(mass eigenstate) dark pions.

It is clear from 
Eqs.~(\ref{eq:darkfermionmasseigenstates})
that a non-zero Yukawa coupling
necessarily splits the fermion masses, and thus 
there is always some (possibly small) mass hierarchy
between $\pi_1$, $K$, and $\pi_2$ (and $\eta$), 
see Eqs.~(\ref{eq:mpi1})--(\ref{eq:mpi2}).
While it is straightforward to calculate the
couplings of all of the dark pions to the Standard Model,
here we focus only on the lightest pions.  
For instance, strong decays of 
$\pi_{\rm heavy}, K \rightarrow \pi_{\rm light} + X$
are expected to be rapid so long as the dark pion
mass differences are large enough that phase space
does not severely limit their rates. 

The two-pion interactions with the SM gauge sector
take the form
\begin{eqnarray}
W^{\mu,\mp} \left( \pi^\pm_{\rm phys} \partial_\mu \pi^0_{\rm phys} 
                   - \pi^0_{\rm phys} \partial_\mu \pi^\pm_{\rm phys} \right): 
&=& g \, \frac{1 + c_m}{2} \, .
\end{eqnarray}
Several limits are interesting.  First, for $\Delta > 0$
and $\Delta \gg y v$, then $c_m \simeq 1$, and so 
the coupling of the dark pions to the gauge bosons becomes 
$\simeq g$ --  exactly the coupling expected for three 
$SU(2)_L$-triplets to interact via the $SU(2)$ anti-symmetric
tensor contraction.  This is not surprising -- in this limit
the lightest pions are a nearly exactly an $SU(2)_L$ triplet
with only $(y v)/\Delta$-suppressed mixings into the
other dark pions.  

Next consider $\Delta < 0$, while still $|\Delta| \gg y v$.
Now $c_m \simeq -1$, and the coupling of the dark pions to the 
gauge bosons becomes $\simeq 0$.  This is again unsurprising --
in this limit the lightest pions are a nearly exact $SU(2)_R$
triplet that does not couple with $SU(2)_L$ gauge bosons.

Finally, when $\Delta \ll y v$ (and thus $c_m \simeq 0$)
the splittings among the dark
pions are dominated by electroweak symmetry breaking contributions.
In this case, the would-be $SU(2)_L$ triplet and $SU(2)_R$ triplets
are fully mixed, and each share an approximately $g/2$ coupling 
to $SU(2)_L$ gauge bosons.

Single pion interactions with one gauge boson and one Higgs boson
are the most interesting (and most relevant for pion decay).
We obtain:
\begin{eqnarray}
(\pi^\pm_{\rm phys} \partial_\mu h - h \partial_\mu \pi^\pm_{\rm phys})  W^{\mu,\mp}: 
& &  \frac{M_W}{v} \, \times \, 
  \left( \sqrt{2} c_D \epsilon y s_m \frac{4 \pi f^2}{m_K^2} \right) 
  \, \times \, \left( \frac{m_h^2}{m_K^2 - m_h^2} \right) \nonumber \\ 
  (\pi^0_{\rm phys} \partial_\mu h - h \partial_\mu \pi^0_{\rm phys}) Z^\mu: 
& & \frac{M_Z}{v} \, \times \, 
  \left( \sqrt{2} c_D \epsilon y s_m \frac{4 \pi f^2}{m_K^2} \right) 
  \, \times \, \left( \frac{m_h^2}{m_K^2 - m_h^2} \right) \nonumber \\ 
  \pi^\pm_{\rm phys} \bar{f}' f: 
& & \sqrt{2} \, 
  \left( \frac{m_{f'}}{v} P_L - \frac{m_f}{v} P_R \right) (2 T_3^f) 
  \, \times \, 
  \left( \sqrt{2} c_D \epsilon y s_m \frac{4 \pi f^2}{m_K^2} \right) 
  \nonumber \\ 
\pi^0_{\rm phys} \bar{f} f: 
& & i \, \left( \frac{m_f}{v} \gamma_5 \right) (2 T_3^f) \, \times \, 
  \left( \sqrt{2} c_D \epsilon y s_m \frac{4 \pi f^2}{m_K^2} \right) 
\label{eq:fourflavorcouplings}
\end{eqnarray} 
From these expressions, we can identify 
\begin{equation} 
\frac{1}{v_\pi} = \frac{1}{v} \, \times \, 
  \left( \sqrt{2} c_D \epsilon y s_m \frac{4 \pi f^2}{m_K^2} \right) 
  \; , \qquad \xi = \frac{m_h^2}{m_K^2 - m_h^2} 
\end{equation}
This is the main result for the four-flavor theory.
We find that the pion interactions with the gauge bosons and Higgs
boson are suppressed relative to the fermion couplings by 
a factor $m_h^2/(m_K^2 - m_h^2)$ that becomes roughly $m_h^2/m_K^2$
for larger dark kaon masses.  This relative suppression in gauge/Higgs boson
couplings to the fermion couplings is exactly what happened in
the two-flavor, custodially-symmetric model.  

The four-flavor model is, essentially, one ultraviolet completion 
of the two-flavor theory with higher-dimensional operators that 
are both custodially symmetric and minimal flavor violating.
The dimension-7 operators that lead to interactions with the fermions
are matched at $\Lambda^3 = 4 \pi f m_K^2$; 
the dimension-9 operator that leads to the interactions with the
gauge bosons and Higgs boson is matched at $\Lambda^5 = 4 \pi f^3 m_K^2$; 
with the coefficient $c_{9C} \propto \lambda_h$ the quartic coupling 
of the Higgs sector.

\section{Dark Sector Custodial Violation}
\label{sec:darksectorcustodialviolation}

We have focused on dark sectors that preserve custodial $SU(2)$.
In practice this means that renormalizable and higher dimensional
operators involving dark fermions do not involve explicit 
$t^3_R$ -- this only appears from the custodially violating 
SM spurions proportional to $g' Y$ or $\mathcal{Y}^{\slashed{C}}_{ij}$). 

Naturally, it is interesting to consider what happens when
explicit $t^3_R$ is introduced.  In the $SU(2)_R$ model,
this is possible already at the renormalizable level.
One can include $M' F t^3_R \hat{F}$ in addition to
$M F \hat{F}$.  This is equivalent to simply writing
different dark fermion masses for the $Y = +1/2$ and
$Y = -1/2$ states under $U(1)_Y$.  

In the $SU(2)_L$ model, gauge invariance forbids a dimension-3
term violating custodial $SU(2)$.  At dimension-5 there is an
interaction:
\begin{eqnarray}
\mathscr{L} &=& 
c_{5V}  \frac{(F_1 t^a_L F_2) 
        \mathrm{Tr} \, \mathcal{H}^\dagger t^a_L \mathcal{H} t^3_R}{\Lambda}
\label{eq:dim5triplet}
\end{eqnarray}
that violates custodial $SU(2)$.  With two Higgs bifundamentals,
the group contractions are
\begin{eqnarray}
(\mathbf{2}_L, \mathbf{2}_R) \otimes (\mathbf{2}_L, \mathbf{2}_R)  
&=& (\mathbf{1}_L, \mathbf{1}_R) \oplus (\mathbf{3}_L, \mathbf{3}_R)  
\end{eqnarray} 
where the surviving combinations are precisely those in 
Eqs.~(\ref{eq:dim5singlet}),(\ref{eq:dim5triplet}).
(The would-be $(\mathbf{3}_L,\mathbf{1}_R)$ or 
$(\mathbf{1}_L,\mathbf{3}_R)$ involves
$\mathrm{Tr} \, \mathcal{H}^\dagger t^a_{L,R} \mathcal{H}$ 
that simply vanishes.) 
The only way we can write a gauge-invariant term of the 
form Eq.~(\ref{eq:dim5triplet}) is to use $t^3_R$ of $SU(2)_R$, 
and hence is custodially violating.  

The low energy effective theory including higher dimensional operators
up to $\mathcal{O}(v^2/\Lambda)$ 
can again be described by a non-linear sigma model,
\begin{eqnarray}
\mathscr{L} &=& \frac{f^2}{4} \mathrm{Tr} \, 
                 (D_\mu \Sigma)^\dagger D^\mu \Sigma \, 
                 + \, 4 \pi f^3 \left( M + c_{5M} \frac{v^2}{\Lambda} \right) 
                 \mathrm{Tr} \, (\Sigma^\dagger + h.c.)  \nonumber \\
& &{} + \, c_{5V} \frac{4 \pi f^3}{\Lambda} 
        \mathrm{Tr} (\Sigma_L t^a_L)
        \mathrm{Tr} \, (\mathcal{H}^\dagger t^a_L \mathcal{H} t^3_R + h.c.) 
        \label{eq:nlsmcustodialviolating}
\end{eqnarray}
Expanding the non-linear sigma model up to $\mathcal{O}( \pi^2 )$, 
we obtain
\begin{eqnarray}
\mathscr{L} &=& \mathrm{Tr} \, D_\mu \pi^a D^\mu \pi^a 
                - \frac{1}{2} m_\pi^2 \pi^a \pi^a
                - c_{5V} \frac{4 \pi f^2}{\Lambda} H^\dagger \pi^a t^a_L H
\label{eq:crappytriplet}
\end{eqnarray}
where $m_\pi^2 = 4 \pi f (M + \mathcal{O}(v^2/\Lambda))$,
and we have written the single pion -- Higgs interaction in the
more familiar form using Higgs doublet notation.  
This Lagrangian is precisely that of a 
``crappy triplet model''\footnote{``Crappy'' in the sense
that the linear term for $\pi^a$ causes it to acquire a 
custodially-violating vev.}, e.g. \cite{SekharChivukula:2007gi}. 
That is, the two-flavor $SU(2)_L$ dark sector with a 
dimension-5 custodially-violating interaction with the 
SM Higgs sector provides an ultraviolet completion 
of the SM extended to include a real triplet.
Higher order terms in the chiral Lagrangian lead to the
usual pion self-interactions as well as interactions of 
multiple pions with Higgs fields.  

In this theory, we see that the dark pion interactions
with the SM arise at a comparatively low dimension operator,
Eq.~(\ref{eq:dim5triplet}).  The explicit custodial violation 
causes the dark pions to acquire a ``triplet'' vev 
\begin{eqnarray}
v_T \equiv \langle \pi^a \rangle &\sim& c_{5V} \frac{f v^2}{\Lambda M} \, .
\end{eqnarray}
Obviously this is highly constrained by electroweak 
precision data.  Nevertheless, following Ref.~\cite{SekharChivukula:2007gi}
one can proceed as usual, shift to the new vacuum,
and extract the effective interactions from 
Eq.~(\ref{eq:crappytriplet}).  
The result is that there is a neutral singlet that mixes
with the Higgs boson and a charged scalar that mixes
with the charged Higgs Goldstones.  Diagonalizing these
interactions leads to 
\begin{eqnarray}
\left( 
\begin{array}{c}
G^\pm_{\rm phys} \\
\pi^\pm_{\rm phys}
\end{array}
\right) &=& 
\left( 
\begin{array}{cc}
\cos\delta  & \sin\delta \\
-\sin\delta & \cos\delta 
\end{array}
\right) 
\left( 
\begin{array}{c}
G^\pm \\
\pi^\pm
\end{array}
\right) 
\end{eqnarray}
where $\sin\delta \sim v_T/\sqrt{v^2 + v_T^2}$. 
The 
interactions of the charged dark pions are obtained
by replacing $G^+$ with $\pi_{\rm phys}^\pm$.
Just like in the two-flavor chiral model, 
this leads to gaugephilic branching ratios.
However, unlike the two-flavor chiral model, 
there is no neutral dark pion / neutral Goldstone mixing.  

\section{Discussion}
\label{sec:discussion}

In this paper we have studied dark sectors that arise from
a new, strongly-coupled confining gauge group $SU(N_D)$ with dark fermions
transforming under the electroweak part of the SM\@.
In dark sectors that preserve custodial $SU(2)$ in their
interactions with the SM, a custodial triplet of
dark pions appears in the low energy effective theory.
The low energy effective interactions with the SM
can be classified by the custodial symmetry, 
leading to two distinct possibilities:
``Gaugephilic'': where $\pi_D^0 \ra Z h$, $\pi_D^\pm \rightarrow W h$
dominate once kinematically open, and 
``Gaugephobic'': where $\pi_D^0 \ra \bar{f} f$, $\pi_D^\pm \rightarrow \bar{f}' f$
dominate.  These classifications assume the only
sources of custodial $SU(2)$ breaking are from 
the SM:  the gauging of hypercharge, $g' Y$, and the difference 
between the up-type and down-type Yukawa couplings,
$\mathcal{Y}^{\slashed{C}}_{ij}$.

The simplest theories that exhibited the gaugephobic and gaugephilic
classifications contained two-flavors, and we examined one chiral theory and two vector-like theories. The chiral theory is familiar from bosonic technicolor/strongly-coupled induced electroweak symmetry breaking. There, the dominant source of dark pion interactions with the SM is from Goldstone-pion mixing and leads to a gaugephilic decay pattern. In the vector-like theories, dark pion interactions with the SM arise through higher dimensional operators. If we demand custodial $SU(2)$ invariance in these higher dimensional operators, we find that interactions between the $\pi_D$ and gauge bosons first occur at dimension-9 (in the UV) while $\pi_D \bar f f$ operators can be written at dimension-7. The mismatch in operator dimension means the vector-like theories are gaugephobic. 

Next, we examined a four-flavor theory. With the proper electroweak charge assignment, this scenario can have both vector-like and chiral masses among its dark fermions, and is therefore a hybrid of the chiral and vector scenarios. The most phenomenologically interesting limit is when the chiral mass is small compared to the vector-like mass. In this case, we find  the lightest custodial $SU(2)$ triplet of dark pions 
have gaugephobic interactions with the SM in which $\pi^0 \ra Z h$, $\pi^\pm \rightarrow W h$ are suppressed by $\simeq m_h^2/m_K^2$ relative to fermionic decays. In the chiral lagrangian for the full multiplet of $15$ dark pions, this arises 
through a cancellation between the dark pion mixing with the 
Goldstones of the SM and dark pion mixing with the Higgs 
boson of the SM\@.  Decoupling the heavier dark pion multiplets such that only the lightest triplet remains, the four-flavor theory maps into a two-flavor theory with higher dimensional operators
that preserve custodial $SU(2)$ and are minimally flavor violating.
The custodial $SU(2)$ symmetry of these interactions 
automatically leads to the operator suppression
$\simeq v^2/\Lambda^2$, in agreement with what we found
by explicit calculation of the four-flavor theory.

In theories that preserve custodial $SU(2)$, the 
neutral dark pion decays to the SM through ``gaugephobic''
or ``gaugephilic'' interactions with a suppressed
rate of $\pi^0 \ra \gamma\gamma$.  In each of the theories
considered, there is no axial anomaly contribution 
to the decay.  However, since the dark pions do have
interactions with SM, and the SM fermions have an
anomalous axial-vector current, the decay
$\pi^0 \ra \gamma\gamma$ does occur, but is suppressed
by the same $1/v_\pi$ that suppresses the direct decay
$\pi^0 \ra {\rm SM} \, {\rm SM}$.  In the Standard Model,
the analogy would be to imagine that the up and down quarks
have an exact custodial $SU(2)$ symmetry, i.e., 
$Q_u = - Q_d = 1/2$ and $y_u = y_d$.  In this case, 
the anomaly contribution to $\pi^0 \ra \gamma\gamma$
in the Standard Model would vanish.  However, 
even without the anomaly, the SM $\pi^0$ decays through
the mode $\pi^0 \ra e^+e^-$ proportional to the electron
Yukawa coupling.  This interaction has the same form as 
the two-flavor chiral theory we considered in this paper.
Now there remains a one-loop suppressed contribution 
$\pi^0 \ra \gamma\gamma$ through the electron Yukawa coupling,
but this is highly suppressed compared with the fermionic decay,
which is precisely what happens with the $\pi^0$ of the 
custodial $SU(2)$ symmetric dark sector theories that we have
considered in this paper.

Finally, the astute reader may have noticed that all of
the vector-like dark sector theories with custodially
symmetric interactions with the SM were gaugephobic.
The only gaugephilic case presented in the paper
is the two-flavor chiral theory, which might give
the reader the impression that vector-like theories
are automatically gaugephobic.  This is not the case.
As an explicit counter-example, the custodial triplets in 
Georgi-Machaeck models have gaugephilic couplings 
(e.g.~\cite{Gunion:1989we}).  It will come as no surprise
that we have already constructed strongly-coupled models based
on coset theories that generate the scalar sector 
of Georgi-Machaeck theories as dark pions with
gaugephilic couplings with the SM\@.  The details
will be presented in Ref.~\cite{usgm}.

\section*{Acknowledgments}

We thank S.~Chang, E.~Neil, and B.~Ostdiek for helpful 
discussions as this work was being completed.  
GDK and AM thank the Universities Research Association for 
travel support, and Fermilab for hospitality, where part of 
this work was completed.
This work of GDK and TT was supported in part by the 
U.S. Department of Energy under Grant Number DE-SC0011640.
The work of AM was supported in part by  
the National Science Foundation  
under Grant Numbers PHY-1520966 and PHY-1820860. 

\appendix

\renewcommand{\theequation}{\thesection.\arabic{equation}}

\section{Gaugephobic 2HDMs}
\label{app:2HDM}

We review the application of the $(\mathbf{2},\mathbf{2})$ 
custodial symmetry formalism in the context of general 
two-Higgs doublet models 
(2HDM) \cite{Nagel:2004sw,Maniatis:2006fs,Grzadkowski:2010dj,Haber:2010bw}.
We'll focus on a general CP-conserving 2HDM\@.   

The most general 2HDM potential can be written as
\cite{Haber:1993an,Gunion:2002zf}
\begin{eqnarray}
V_{2HDM} &=& \phantom{+}
m_{11}^2 (\phi_1^\dagger \phi_1) +
m_{22}^2 (\phi_2^\dagger \phi_2) \nonumber
\\ \nonumber
& &
- m_{12}^2 (\phi_1^\dagger \phi_2) -
(m_{12}^2)^* (\phi_2^\dagger \phi_1)
\\ \nonumber
& &
+\frac{1}{2} \lambda_1 (\phi_1^\dagger \phi_1)^2 
+ \frac{1}{2} \lambda_2 (\phi_2^\dagger \phi_2)^2 
+ \lambda_3 (\phi_1^\dagger \phi_1)(\phi_2^\dagger \phi_2)
\\ \nonumber
& &
+ \lambda_4 (\phi_1^\dagger \phi_2)(\phi_2^\dagger \phi_1)
+ \frac{1}{2} [\lambda_5 (\phi_1^\dagger \phi_2)^2 + \lambda_5^* 
(\phi_2^\dagger \phi_1)^2]
\\ \nonumber
& &
+ [\lambda_6 (\phi_1^\dagger \phi_2) + \lambda_6^* 
(\phi_2^\dagger \phi_1)] (\phi_1^\dagger \phi_1) \\ 
& &
+ [\lambda_7 (\phi_1^\dagger 
\phi_2) + \lambda_7^* (\phi_2^\dagger \phi_1)] (\phi_2^\dagger \phi_2)
\label{eq:V2HDM}
\end{eqnarray}
where $m_{11}^2$, $m_{22}^2$, $\lambda_{1,2,3,4}$ are real parameters 
and $m_{12}^2$, $\lambda_{5,6,7}$ complex. 
And $\phi_1$ and $\phi_2$ are two complex scalar doublets 
\begin{eqnarray}
\phi_1 = 
\begin{pmatrix}
\phi_1^+ \\ \phi_1^0
\end{pmatrix},
\qquad
\phi_2 = 
\begin{pmatrix}
\phi_2^+ \\ \phi_2^0
\end{pmatrix},
\end{eqnarray}

In general, $m_{11}^2$, $m_{22}^2$, and $\lambda_{1,2,3,4}$ are real parameters 
while $m_{12}^2$ and $\lambda_{5,6,7}$ can be complex. 
Nevertheless, in this study we restrict our discussion to CP-conserving models, 
by assuming all the parameters of $V_{2HDM}$ are real \cite{Gunion:2002zf}. 
And we also assume the parameters are chosen to 
make $V_{2HDM}$ bounded below so that each of 
the $\phi_i$ acquires a VEV, denoted as $v_1$ and $v_2$ which satisfy 
\begin{eqnarray} 
v_1^2 + v_2^2 = v^2 = (246 \gev)^2 
\label{v246} 
\end{eqnarray} 
and we define 
\begin{eqnarray} 
\tb \equiv \tanb \equiv \frac{v_2}{v_1} 
\end{eqnarray} 

The goal of this section is to demonstrate explicitly that 
it's possible to write a general 2HDM potential in terms of 
a $(\mathbf{2},\mathbf{2})$ custodial symmetry formalism, 
by introducing matrices $M_{ij}$ similar to Eq.~(\ref{eq:h2x2}) 
\begin{eqnarray}
M_{ij} \equiv (\tp_i,\phi_j)=
\begin{pmatrix}
\phi^{0\star}_i & \phi^+_j \\ -\phi^-_i & \phi^0_j
\end{pmatrix} 
\label{eq:M}
\end{eqnarray}
where $i,j=1,2$

It is crucial to our approach that we define the following 
$\tvec{K}$-terms \cite{Nagel:2004sw,Maniatis:2006fs,Grzadkowski:2010dj}
\begin{eqnarray}
\tvec{K} = \left(
\begin{array}{c}
K_0 \\
K_1 \\
K_2 \\
K_3
\end{array}
\right) 
= \left(
\begin{array}{c}
\phi_1^\dagger \phi_1 + \phi_2^\dagger \phi_2 \\
\phi_1^\dagger \phi_2 + \phi_2^\dagger \phi_1 \\
i (\phi_2^\dagger \phi_1 - \phi_1^\dagger \phi_2) \\
\phi_1^\dagger \phi_1 - \phi_2^\dagger \phi_2
\end{array}
\right)
\label{eq:K}
\end{eqnarray}

Given Eqs.~(\ref{eq:M}-\ref{eq:K}), we may write $\tvec{K}$ 
in two different ways, with either $M_{11}$ and $M_{22}$, or $M_{21}$ alone 
\begin{equation}
\begin{array}{rcll}
K_0 &=& \frac12 \tr (M_{11}^\dagger M_{11}+M_{22}^\dagger M_{22}) 
        & = \tr (M_{21}^\dagger M_{21}) \\
K_1 &=& \tr (M_{11}^\dagger M_{22}) 
        & = 2 \re (\det M_{21}^\dagger) \\
K_2 &=& (-i) \tr (M_{11} \tau_3 M_{22}^\dagger) 
        & = -2 \im (\det M_{21}) \\
K_3 &=& \frac12 \tr (M_{11}^\dagger M_{11}-M_{22}^\dagger M_{22}) 
        & = -\tr (M_{21}\tau_3 M_{21}^\dagger) 
\end{array}
\label{eq:K-M}
\end{equation}
Then it is straightforward to verify that $V_{2HDM}$ 
can be written in terms of $\tvec{K}$ in a compact form of 
\begin{eqnarray}
V_{2HDM} = \tvec{\xi}^\trans \tvec{K} + \tvec{K}^\trans \tvec{E} \tvec{K}
\end{eqnarray}
where the mass parameter vector $\tvec{\xi}$ and the 
coupling parameter matrix $\tvec{E}$ are \cite{Grzadkowski:2010dj} 
\begin{eqnarray}
\tvec{\xi} = 
\left(
\begin{array}{c}
\frac{1}{2} (m_{11}^2+m_{22}^2) \\
- \re(m_{12}^2) \\
\im(m_{12}^2) \\
\frac{1}{2} (m_{11}^2-m_{22}^2)
\end{array}
\right)
\label{eq:xi}
\end{eqnarray}
\begin{eqnarray}
\tvec{E} = \frac{1}{4}
\left(
\begin{array}{cccc}
\frac{1}{2} (\lambda_1 + \lambda_2) + \lambda_3 & 
\re(\lambda_6 + \lambda_7) & 
- \im(\lambda_6 + \lambda_7) & 
\frac{1}{2} (\lambda_1 - \lambda_2) \\
\re(\lambda_6 + \lambda_7) & 
\lambda_4 + \re(\lambda_5) & 
- \im(\lambda_5) & 
\re(\lambda_6-\lambda_7) \\
- \im(\lambda_6 + \lambda_7) & 
-\im(\lambda_5) & 
\lambda_4 - \re(\lambda_5) & 
- \im(\lambda_6-\lambda_7) \\
\frac{1}{2} (\lambda_1 - \lambda_2) & 
\re(\lambda_6-\lambda_7) & 
- \im(\lambda_6 -\lambda_7) & 
\frac{1}{2}(\lambda_1 + \lambda_2) - \lambda_3
\end{array}
\right) 
\label{eq:E}
\end{eqnarray}
As a consequence of Eq.~(\ref{eq:K-M}), there are actually 
two types of custodial transformations to the potential \cite{Pomarol:1993mu}:
Type I: 
$M_{11}$ and $M_{22}$ transform as 
\begin{eqnarray}
M_{ii} \lra L M_{ii} R^\dagger \for  i=1,2 
\end{eqnarray}
where $L$ and $R$ are $SU(2)_L$ and $SU(2)_R$ matrices. 
Type II: 
In this case, it's $M_{21}$ which transforms as 
\begin{eqnarray}
M_{21} \lra L M_{21} R^\dagger 
\end{eqnarray}
The potential $V_{2HDM}$ preserves custodial symmetry 
if it is invariant under either type of the custodial transformations. 

Nevertheless, recall that there is an explicit $\tau_3$ in Eq.~(\ref{eq:K-M}). 
In fact, it's a $(\tau_3)_R$ which appears either in the $K_2$ term 
under the Type I custodial transformation, or in the $K_3$ term for the Type II. 
Since $(\tau_3)_R$ breaks custodial symmetry explicitly, 
$K_2$ term should be absent from $V_{2HDM}$ with Type I 
custodial symmetry, same as $K_3$ term for Type II. 
Apparently, to meet this requirement the corresponding entries 
in $\tvec{\xi}$ and $\tvec{E}$ must vanish. 

With the argument above, the conditions for a custodial symmetric 
2HDM potential can be summarized as: 

Type I: 
\begin{eqnarray} 
\tvec{\xi}_{I} = 
\begin{pmatrix} \cdot \\ \cdot \\ 0 \\ \cdot \end{pmatrix}, \quad \quad 
\tvec{E}_{I} = 
\begin{pmatrix} 
\cdot & \cdot & 0 & \cdot \\ 
\cdot & \cdot & 0 & \cdot \\ 
0 & 0 &  0  & 0 \\ 
\cdot & \cdot & 0 & \cdot 
\end{pmatrix} 
\label{CT-I} 
\end{eqnarray} 

Type II: 
\begin{eqnarray} 
\tvec{\xi}_{II} = 
\begin{pmatrix} \cdot \\ \cdot \\ \cdot \\ 0 \end{pmatrix}, \quad \quad 
\tvec{E}_{II} = 
\begin{pmatrix} 
\cdot & \cdot & \cdot & 0 \\ 
\cdot & \cdot & \cdot & 0 \\ 
\cdot & \cdot & \cdot & 0 \\ 
0 & 0 &  0  & 0 
\end{pmatrix} 
\label{CT-II} 
\end{eqnarray} 
For a $CP$-conserving 2HDM: 
\begin{eqnarray}
\tvec{\xi}_{CP} = 
\left(
\begin{array}{c}
\frac{1}{2} (m_{11}^2+m_{22}^2) \\
- m_{12}^2 \\
0 \\
\frac{1}{2} (m_{11}^2-m_{22}^2)
\end{array}
\right)
\label{eq:xi_CP}
\end{eqnarray}
\begin{eqnarray}
\tvec{E}_{CP} = \frac{1}{4}
\left(
\begin{array}{cccc}
\frac{1}{2} (\lambda_1 + \lambda_2) + \lambda_3 & 
\lambda_6 + \lambda_7 & 
0 & 
\frac{1}{2} (\lambda_1 - \lambda_2) \\
\lambda_6 + \lambda_7 & 
\lambda_4 + \lambda_5 & 
0 & 
\lambda_6-\lambda_7 \\
0 & 
0 & 
\lambda_4 - \lambda_5 & 
0 \\
\frac{1}{2} (\lambda_1 - \lambda_2) & 
\lambda_6-\lambda_7 & 
0 & 
\frac{1}{2}(\lambda_1 + \lambda_2) - \lambda_3
\end{array}
\right) 
\label{eq:E_CP}
\end{eqnarray}

Compare Eqs.~(\ref{eq:xi_CP}-\ref{eq:E_CP}) to (\ref{CT-I}), 
we see that to preserve Type-I custodial symmetry, 
the condition required is 
\begin{eqnarray}
\lambda_4 = \lambda_5 
\label{eq:l45} 
\end{eqnarray}
Similarly, the conditions for a Type-II custodial symmetry are 
\begin{equation}
\begin{array}{rcl}
m_{11}^2  &=& m_{22}^2 \\ 
\lambda_1 &=& \lambda_2 \\ 
\lambda_6 &=& \lambda_7 \\ 
\lambda_3 &=& \frac{1}{2} (\lambda_1 + \lambda_2) = \lambda_1 
\end{array}
\label{eq:l13} 
\end{equation}

As is well-known, the observable that measures custodial violation
is the $\rho$-parameter.  Assuming the first three conditions 
of Eq.~(\ref{eq:l13}), the one-loop contributions to
$\Delta \rho$ \cite{Gunion:1989we} from either Type-I or Type-II models
can be calculated to the leading order in $v^2$ as
\begin{eqnarray} 
\Delta \rho = \frac{1}{192 \pi^2} \left( \frac{v^2}{m_A^2} \right) 
(\lambda_4 - \lambda_5) (\lambda_1 - \lambda_3) 
\label{eq:delta-rho} 
\end{eqnarray} 
where $m_A$ is the mass of the heavy pseudoscalar Higgs state $A^0$ in 2HDM\@. 
We explicitly see that $\Delta \rho$ is proportional to 
$(\lambda_4 - \lambda_5)$ \emph{and} $(\lambda_1 - \lambda_3)$, 
which can be identified with the contribution from 
Type-I and Type-II, correspondingly. 

We can also map the general Type II 2HDM model onto the 
the minimal supersymmetric model (MSSM) where
the $\lambda_i$ are \cite{Gunion:2002zf}
\begin{equation}
\begin{array}{rcl}
\lambda_1 &=& \lambda_2 = \frac{1}{4} (g^2 + g'^2) \\ 
\lambda_3 &=& \frac{1}{4} (g^2 - g'^2) \\ 
\lambda_4 &=& - \frac{1}{2} g^2 \\ 
\lambda_5 &=& \lambda_6 = \lambda_7 = 0 \, .
\end{array}
\end{equation}
The contribution to $\Delta \rho$ is then
\begin{eqnarray} 
\Delta \rho = \frac{1}{192 \pi^2} \left( \frac{v^2}{m_A^2} \right) 
\left(- \frac{1}{2} g^2 \right) 
\left( \frac{1}{2} g'^2 \right) \, . 
\end{eqnarray} 
The 2HDM potential of the MSSM contains custodial symmetry violation 
with a small but non-zero correction to the $\rho$-parameter.  
The correct is, nevertheless, proportional to $g'^2$
that is precisely the SM violation of custodial symmetry 
by gauging hypercharge.

Phenomenologically, the heavy Higgs states in a 2HDM 
may decay into SM particles if kinematically allowed. 
Comparing to our study of dark mesons, 
we are particularly interested in the branching fractions of 
the charged Higgs $H^\pm$ and the pseudoscalar $A^0$ 
decaying into SM fermion pairs or gauge boson and Higgs pairs, 
especially in the decoupling limit $m_A \gg v$. 
In this limit, Eq.~(\ref{eq:delta-rho}) indicates that $\Delta \rho$ 
is always suppressed by two powers of the heavy mass scale $m_A$, 
which means the amount of possible custodial symmetry violation 
is restricted to be relatively small. As a result, one can say that 
2HDM becomes custodially symmetric in the decoupling limit. 

As for the decay branching fractions, though the couplings of 
$H^\pm$ and $A^0$ to SM fermions are usually model dependent, 
their values are proportional to $\tanb$ or $\cotb$ \cite{Gunion:1989we} 
\begin{eqnarray}
C_{ff} \propto g \frac{m_f}{m_W} (\tanb ~\, \text{or} ~\cotb) 
\label{eq:cff} 
\end{eqnarray}
On the other hand, the couplings to SM 
gauge bosons and SM Higgs are proportional to $\cosbma$ \cite{Gunion:1989we} 
\begin{eqnarray} 
C_{Wh} \propto g \cosbma 
\label{eq:cwh} 
\end{eqnarray} 
where $\alpha$ is the CP-even scalar mixing angle, and in the decoupling limit, 
\begin{eqnarray} 
\cosbma = {\cal O} \left( \frac{v^2}{m_A^2} \right) 
\label{eq:cba} 
\end{eqnarray} 
Compare Eq.~(\ref{eq:cff}) to Eq.~(\ref{eq:cwh}), we see that to the 
leading order in $v^2$, 
\begin{eqnarray} 
\frac{C_{Wh}}{C_{ff}} \propto \cosbma 
\propto {\cal O} \left( \frac{v^2}{m_A^2} \right) 
\label{eq:2HDMgaugephobic} 
\end{eqnarray} 
Therefore, in the decoupling limit a 2HDM becomes custodially symmetric, 
and the decays of its heavy states to SM particle in this limit 
are gaugephobic.

\bibliographystyle{utphys}
\bibliography{DP}

\providecommand{\href}[2]{#2}\begingroup\raggedright\begin{thebibliography}{100}

\bibitem{Simmons:1988fu}
E.~H. Simmons, ``{Phenomenology of a Technicolor Model With Heavy Scalar
  Doublet},''
\href{http://dx.doi.org/10.1016/0550-3213(89)90296-4}{{\em Nucl. Phys.} {\bf
  B312} (1989)  253--268}.
%%CITATION = NUPHA,B312,253;%%.

\bibitem{Samuel:1990dq}
S.~Samuel, ``{BOSONIC TECHNICOLOR},''
\href{http://dx.doi.org/10.1016/0550-3213(90)90378-Q}{{\em Nucl. Phys.} {\bf
  B347} (1990)  625--650}.
%%CITATION = NUPHA,B347,625;%%.

\bibitem{Dine:1990jd}
M.~Dine, A.~Kagan, and S.~Samuel, ``{Naturalness in Supersymmetry, or Raising
  the Supersymmetry Breaking Scale},''
\href{http://dx.doi.org/10.1016/0370-2693(90)90847-Y}{{\em Phys. Lett.} {\bf
  B243} (1990)  250--256}.
%%CITATION = PHLTA,B243,250;%%.

\bibitem{Kagan:1990az}
A.~Kagan and S.~Samuel, ``{The Family mass hierarchy problem in bosonic
  technicolor},''
\href{http://dx.doi.org/10.1016/0370-2693(90)90492-O}{{\em Phys. Lett.} {\bf
  B252} (1990)  605--610}.
%%CITATION = PHLTA,B252,605;%%.

\bibitem{Kagan:1991gh}
A.~Kagan and S.~Samuel, ``{Renormalization group aspects of bosonic
  technicolor},''
\href{http://dx.doi.org/10.1016/0370-2693(91)91535-4}{{\em Phys. Lett.} {\bf
  B270} (1991)  37--44}.
%%CITATION = PHLTA,B270,37;%%.

\bibitem{Carone:1992rh}
C.~D. Carone and E.~H. Simmons, ``{Oblique corrections in technicolor with a
  scalar},'' \href{http://dx.doi.org/10.1016/0550-3213(93)90187-T}{{\em Nucl.
  Phys.} {\bf B397} (1993)  591--615},
\href{http://arxiv.org/abs/hep-ph/9207273}{{\tt arXiv:hep-ph/9207273
  [hep-ph]}}.
%%CITATION = HEP-PH/9207273;%%.

\bibitem{Carone:1993xc}
C.~D. Carone and H.~Georgi, ``{Technicolor with a massless scalar doublet},''
  \href{http://dx.doi.org/10.1103/PhysRevD.49.1427}{{\em Phys. Rev.} {\bf D49}
  (1994)  1427--1436},
\href{http://arxiv.org/abs/hep-ph/9308205}{{\tt arXiv:hep-ph/9308205
  [hep-ph]}}.
%%CITATION = HEP-PH/9308205;%%.

\bibitem{Dobrescu:1997kt}
B.~A. Dobrescu and J.~Terning, ``{Negative contributions to S in an effective
  field theory},'' \href{http://dx.doi.org/10.1016/S0370-2693(97)01304-X}{{\em
  Phys. Lett.} {\bf B416} (1998)  129--136},
\href{http://arxiv.org/abs/hep-ph/9709297}{{\tt arXiv:hep-ph/9709297
  [hep-ph]}}.
%%CITATION = HEP-PH/9709297;%%.

\bibitem{Antola:2009wq}
M.~Antola, M.~Heikinheimo, F.~Sannino, and K.~Tuominen, ``{Unnatural Origin of
  Fermion Masses for Technicolor},''
  \href{http://dx.doi.org/10.1007/JHEP03(2010)050}{{\em JHEP} {\bf 03} (2010)
  050},
\href{http://arxiv.org/abs/0910.3681}{{\tt arXiv:0910.3681 [hep-ph]}}.
%%CITATION = ARXIV:0910.3681;%%.

\bibitem{Azatov:2011ht}
A.~Azatov, J.~Galloway, and M.~A. Luty, ``{Superconformal Technicolor},''
  \href{http://dx.doi.org/10.1103/PhysRevLett.108.041802}{{\em Phys. Rev.
  Lett.} {\bf 108} (2012)  041802},
\href{http://arxiv.org/abs/1106.3346}{{\tt arXiv:1106.3346 [hep-ph]}}.
%%CITATION = ARXIV:1106.3346;%%.

\bibitem{Azatov:2011ps}
A.~Azatov, J.~Galloway, and M.~A. Luty, ``{Superconformal Technicolor: Models
  and Phenomenology},''
  \href{http://dx.doi.org/10.1103/PhysRevD.85.015018}{{\em Phys. Rev.} {\bf
  D85} (2012)  015018},
\href{http://arxiv.org/abs/1106.4815}{{\tt arXiv:1106.4815 [hep-ph]}}.
%%CITATION = ARXIV:1106.4815;%%.

\bibitem{Gherghetta:2011na}
T.~Gherghetta and A.~Pomarol, ``{A Distorted MSSM Higgs Sector from Low-Scale
  Strong Dynamics},'' \href{http://dx.doi.org/10.1007/JHEP12(2011)069}{{\em
  JHEP} {\bf 12} (2011)  069},
\href{http://arxiv.org/abs/1107.4697}{{\tt arXiv:1107.4697 [hep-ph]}}.
%%CITATION = ARXIV:1107.4697;%%.

\bibitem{Galloway:2013dma}
J.~Galloway, M.~A. Luty, Y.~Tsai, and Y.~Zhao, ``{Induced Electroweak Symmetry
  Breaking and Supersymmetric Naturalness},''
  \href{http://dx.doi.org/10.1103/PhysRevD.89.075003}{{\em Phys. Rev.} {\bf
  D89} (2014) no.~7, 075003},
\href{http://arxiv.org/abs/1306.6354}{{\tt arXiv:1306.6354 [hep-ph]}}.
%%CITATION = ARXIV:1306.6354;%%.

\bibitem{Chang:2014ida}
S.~Chang, J.~Galloway, M.~Luty, E.~Salvioni, and Y.~Tsai, ``{Phenomenology of
  Induced Electroweak Symmetry Breaking},''
  \href{http://dx.doi.org/10.1007/JHEP03(2015)017}{{\em JHEP} {\bf 03} (2015)
  017},
\href{http://arxiv.org/abs/1411.6023}{{\tt arXiv:1411.6023 [hep-ph]}}.
%%CITATION = ARXIV:1411.6023;%%.

\bibitem{Beauchesne:2015lva}
H.~Beauchesne, K.~Earl, and T.~Gregoire, ``{The spontaneous Z2 breaking Twin
  Higgs},'' \href{http://dx.doi.org/10.1007/JHEP01(2016)130}{{\em JHEP} {\bf
  01} (2016)  130},
\href{http://arxiv.org/abs/1510.06069}{{\tt arXiv:1510.06069 [hep-ph]}}.
%%CITATION = ARXIV:1510.06069;%%.

\bibitem{Harnik:2016koz}
R.~Harnik, K.~Howe, and J.~Kearney, ``{Tadpole-Induced Electroweak Symmetry
  Breaking and pNGB Higgs Models},''
  \href{http://dx.doi.org/10.1007/JHEP03(2017)111}{{\em JHEP} {\bf 03} (2017)
  111},
\href{http://arxiv.org/abs/1603.03772}{{\tt arXiv:1603.03772 [hep-ph]}}.
%%CITATION = ARXIV:1603.03772;%%.

\bibitem{Alanne:2016rpe}
T.~Alanne, M.~T. Frandsen, and D.~Buarque~Franzosi, ``{Testing a dynamical
  origin of Standard Model fermion masses},''
  \href{http://dx.doi.org/10.1103/PhysRevD.94.071703}{{\em Phys. Rev.} {\bf
  D94} (2016)  071703},
\href{http://arxiv.org/abs/1607.01440}{{\tt arXiv:1607.01440 [hep-ph]}}.
%%CITATION = ARXIV:1607.01440;%%.

\bibitem{Galloway:2016fuo}
J.~Galloway, A.~L. Kagan, and A.~Martin, ``{A UV complete partially
  composite-pNGB Higgs},''
  \href{http://dx.doi.org/10.1103/PhysRevD.95.035038}{{\em Phys. Rev.} {\bf
  D95} (2017) no.~3, 035038},
\href{http://arxiv.org/abs/1609.05883}{{\tt arXiv:1609.05883 [hep-ph]}}.
%%CITATION = ARXIV:1609.05883;%%.

\bibitem{Agugliaro:2016clv}
A.~Agugliaro, O.~Antipin, D.~Becciolini, S.~De~Curtis, and M.~Redi, ``{UV
  complete composite Higgs models},''
  \href{http://dx.doi.org/10.1103/PhysRevD.95.035019}{{\em Phys. Rev.} {\bf
  D95} (2017) no.~3, 035019},
\href{http://arxiv.org/abs/1609.07122}{{\tt arXiv:1609.07122 [hep-ph]}}.
%%CITATION = ARXIV:1609.07122;%%.

\bibitem{Barducci:2018yer}
D.~Barducci, S.~De~Curtis, M.~Redi, and A.~Tesi, ``{An almost elementary Higgs:
  Theory and Practice},'' \href{http://dx.doi.org/10.1007/JHEP08(2018)017}{{\em
  JHEP} {\bf 08} (2018)  017},
\href{http://arxiv.org/abs/1805.12578}{{\tt arXiv:1805.12578 [hep-ph]}}.
%%CITATION = ARXIV:1805.12578;%%.

\bibitem{Bellazzini:2014yua}
B.~Bellazzini, C.~Csáki, and J.~Serra, ``{Composite Higgses},''
  \href{http://dx.doi.org/10.1140/epjc/s10052-014-2766-x}{{\em Eur. Phys. J.}
  {\bf C74} (2014) no.~5, 2766},
\href{http://arxiv.org/abs/1401.2457}{{\tt arXiv:1401.2457 [hep-ph]}}.
%%CITATION = ARXIV:1401.2457;%%.

\bibitem{Graham:2015cka}
P.~W. Graham, D.~E. Kaplan, and S.~Rajendran, ``{Cosmological Relaxation of the
  Electroweak Scale},''
  \href{http://dx.doi.org/10.1103/PhysRevLett.115.221801}{{\em Phys. Rev.
  Lett.} {\bf 115} (2015) no.~22, 221801},
\href{http://arxiv.org/abs/1504.07551}{{\tt arXiv:1504.07551 [hep-ph]}}.
%%CITATION = ARXIV:1504.07551;%%.

\bibitem{Antipin:2015jia}
O.~Antipin and M.~Redi, ``{The Half-composite Two Higgs Doublet Model and the
  Relaxion},'' \href{http://dx.doi.org/10.1007/JHEP12(2015)031}{{\em JHEP} {\bf
  12} (2015)  031},
\href{http://arxiv.org/abs/1508.01112}{{\tt arXiv:1508.01112 [hep-ph]}}.
%%CITATION = ARXIV:1508.01112;%%.

\bibitem{Batell:2017kho}
B.~Batell, M.~A. Fedderke, and L.-T. Wang, ``{Relaxation of the Composite Higgs
  Little Hierarchy},'' \href{http://dx.doi.org/10.1007/JHEP12(2017)139}{{\em
  JHEP} {\bf 12} (2017)  139},
\href{http://arxiv.org/abs/1705.09666}{{\tt arXiv:1705.09666 [hep-ph]}}.
%%CITATION = ARXIV:1705.09666;%%.

\bibitem{Nussinov:1985xr}
S.~Nussinov, ``{TECHNOCOSMOLOGY: COULD A TECHNIBARYON EXCESS PROVIDE A
  'NATURAL' MISSING MASS CANDIDATE?},''
\href{http://dx.doi.org/10.1016/0370-2693(85)90689-6}{{\em Phys.Lett.} {\bf
  B165} (1985)  55}.
%%CITATION = PHLTA,B165,55;%%.

\bibitem{Chivukula:1989qb}
R.~S. Chivukula and T.~P. Walker, ``{TECHNICOLOR COSMOLOGY},''
\href{http://dx.doi.org/10.1016/0550-3213(90)90151-3}{{\em Nucl.Phys.} {\bf
  B329} (1990)  445}.
%%CITATION = NUPHA,B329,445;%%.

\bibitem{Barr:1990ca}
S.~M. Barr, R.~S. Chivukula, and E.~Farhi, ``{Electroweak Fermion Number
  Violation and the Production of Stable Particles in the Early Universe},''
\href{http://dx.doi.org/10.1016/0370-2693(90)91661-T}{{\em Phys.Lett.} {\bf
  B241} (1990)  387--391}.
%%CITATION = PHLTA,B241,387;%%.

\bibitem{Barr:1991qn}
S.~M. Barr, ``{Baryogenesis, sphalerons and the cogeneration of dark matter},''
\href{http://dx.doi.org/10.1103/PhysRevD.44.3062}{{\em Phys.Rev.} {\bf D44}
  (1991)  3062--3066}.
%%CITATION = PHRVA,D44,3062;%%.

\bibitem{Kaplan:1991ah}
D.~B. Kaplan, ``{A Single explanation for both the baryon and dark matter
  densities},''
\href{http://dx.doi.org/10.1103/PhysRevLett.68.741}{{\em Phys.Rev.Lett.} {\bf
  68} (1992)  741--743}.
%%CITATION = PRLTA,68,741;%%.

\bibitem{Chivukula:1992pn}
R.~S. Chivukula, A.~G. Cohen, M.~E. Luke, and M.~J. Savage, ``{A Comment on the
  strong interactions of color - neutral technibaryons},''
  \href{http://dx.doi.org/10.1016/0370-2693(93)91836-C}{{\em Phys.Lett.} {\bf
  B298} (1993)  380--382},
\href{http://arxiv.org/abs/hep-ph/9210274}{{\tt arXiv:hep-ph/9210274
  [hep-ph]}}.
%%CITATION = HEP-PH/9210274;%%.

\bibitem{Bagnasco:1993st}
J.~Bagnasco, M.~Dine, and S.~D. Thomas, ``{Detecting technibaryon dark
  matter},'' \href{http://dx.doi.org/10.1016/0370-2693(94)90830-3}{{\em
  Phys.Lett.} {\bf B320} (1994)  99--104},
\href{http://arxiv.org/abs/hep-ph/9310290}{{\tt arXiv:hep-ph/9310290
  [hep-ph]}}.
%%CITATION = HEP-PH/9310290;%%.

\bibitem{Khlopov:2008ty}
M.~{\relax Yu}. Khlopov and C.~Kouvaris, ``{Composite dark matter from a model
  with composite Higgs boson},''
  \href{http://dx.doi.org/10.1103/PhysRevD.78.065040}{{\em Phys. Rev.} {\bf
  D78} (2008)  065040},
\href{http://arxiv.org/abs/0806.1191}{{\tt arXiv:0806.1191 [astro-ph]}}.
%%CITATION = ARXIV:0806.1191;%%.

\bibitem{Ryttov:2008xe}
T.~A. Ryttov and F.~Sannino, ``{Ultra Minimal Technicolor and its Dark Matter
  TIMP},'' \href{http://dx.doi.org/10.1103/PhysRevD.78.115010}{{\em Phys. Rev.}
  {\bf D78} (2008)  115010},
\href{http://arxiv.org/abs/0809.0713}{{\tt arXiv:0809.0713 [hep-ph]}}.
%%CITATION = ARXIV:0809.0713;%%.

\bibitem{Hambye:2009fg}
T.~Hambye and M.~H.~G. Tytgat, ``{Confined hidden vector dark matter},''
  \href{http://dx.doi.org/10.1016/j.physletb.2009.11.050}{{\em Phys. Lett.}
  {\bf B683} (2010)  39--41},
\href{http://arxiv.org/abs/0907.1007}{{\tt arXiv:0907.1007 [hep-ph]}}.
%%CITATION = ARXIV:0907.1007;%%.

\bibitem{Bai:2010qg}
Y.~Bai and R.~J. Hill, ``{Weakly Interacting Stable Pions},''
  \href{http://dx.doi.org/10.1103/PhysRevD.82.111701}{{\em Phys.Rev.} {\bf D82}
  (2010)  111701},
\href{http://arxiv.org/abs/1005.0008}{{\tt arXiv:1005.0008 [hep-ph]}}.
%%CITATION = ARXIV:1005.0008;%%.

\bibitem{Lewis:2011zb}
R.~Lewis, C.~Pica, and F.~Sannino, ``{Light Asymmetric Dark Matter on the
  Lattice: SU(2) Technicolor with Two Fundamental Flavors},''
  \href{http://dx.doi.org/10.1103/PhysRevD.85.014504}{{\em Phys.Rev.} {\bf D85}
  (2012)  014504},
\href{http://arxiv.org/abs/1109.3513}{{\tt arXiv:1109.3513 [hep-ph]}}.
%%CITATION = ARXIV:1109.3513;%%.

\bibitem{Frigerio:2012uc}
M.~Frigerio, A.~Pomarol, F.~Riva, and A.~Urbano, ``{Composite Scalar Dark
  Matter},'' \href{http://dx.doi.org/10.1007/JHEP07(2012)015}{{\em JHEP} {\bf
  07} (2012)  015},
\href{http://arxiv.org/abs/1204.2808}{{\tt arXiv:1204.2808 [hep-ph]}}.
%%CITATION = ARXIV:1204.2808;%%.

\bibitem{Buckley:2012ky}
M.~R. Buckley and E.~T. Neil, ``{Thermal Dark Matter from a Confining
  Sector},'' \href{http://dx.doi.org/10.1103/PhysRevD.87.043510}{{\em
  Phys.Rev.} {\bf D87} (2013) no.~4, 043510},
\href{http://arxiv.org/abs/1209.6054}{{\tt arXiv:1209.6054 [hep-ph]}}.
%%CITATION = ARXIV:1209.6054;%%.

\bibitem{Bhattacharya:2013kma}
S.~Bhattacharya, B.~Melić, and J.~Wudka, ``{Pionic Dark Matter},''
  \href{http://dx.doi.org/10.1007/JHEP02(2014)115}{{\em JHEP} {\bf 02} (2014)
  115},
\href{http://arxiv.org/abs/1307.2647}{{\tt arXiv:1307.2647 [hep-ph]}}.
%%CITATION = ARXIV:1307.2647;%%.

\bibitem{Hietanen:2014xca}
A.~Hietanen, R.~Lewis, C.~Pica, and F.~Sannino, ``{Fundamental Composite Higgs
  Dynamics on the Lattice: SU(2) with Two Flavors},''
  \href{http://dx.doi.org/10.1007/JHEP07(2014)116}{{\em JHEP} {\bf 07} (2014)
  116},
\href{http://arxiv.org/abs/1404.2794}{{\tt arXiv:1404.2794 [hep-lat]}}.
%%CITATION = ARXIV:1404.2794;%%.

\bibitem{Marzocca:2014msa}
D.~Marzocca and A.~Urbano, ``{Composite Dark Matter and LHC Interplay},''
  \href{http://dx.doi.org/10.1007/JHEP07(2014)107}{{\em JHEP} {\bf 07} (2014)
  107},
\href{http://arxiv.org/abs/1404.7419}{{\tt arXiv:1404.7419 [hep-ph]}}.
%%CITATION = ARXIV:1404.7419;%%.

\bibitem{Pasechnik:2014ida}
R.~Pasechnik, V.~Beylin, V.~Kuksa, and G.~Vereshkov, ``{Composite scalar Dark
  Matter from vector-like $SU(2)$ confinement},''
  \href{http://dx.doi.org/10.1142/S0217751X16500366}{{\em Int. J. Mod. Phys.}
  {\bf A31} (2016) no.~08, 1650036},
\href{http://arxiv.org/abs/1407.2392}{{\tt arXiv:1407.2392 [hep-ph]}}.
%%CITATION = ARXIV:1407.2392;%%.

\bibitem{Antipin:2014qva}
O.~Antipin, M.~Redi, and A.~Strumia, ``{Dynamical generation of the weak and
  Dark Matter scales from strong interactions},''
  \href{http://dx.doi.org/10.1007/JHEP01(2015)157}{{\em JHEP} {\bf 01} (2015)
  157},
\href{http://arxiv.org/abs/1410.1817}{{\tt arXiv:1410.1817 [hep-ph]}}.
%%CITATION = ARXIV:1410.1817;%%.

\bibitem{Hochberg:2014kqa}
Y.~Hochberg, E.~Kuflik, H.~Murayama, T.~Volansky, and J.~G. Wacker, ``{Model
  for Thermal Relic Dark Matter of Strongly Interacting Massive Particles},''
  \href{http://dx.doi.org/10.1103/PhysRevLett.115.021301}{{\em Phys. Rev.
  Lett.} {\bf 115} (2015) no.~2, 021301},
\href{http://arxiv.org/abs/1411.3727}{{\tt arXiv:1411.3727 [hep-ph]}}.
%%CITATION = ARXIV:1411.3727;%%.

\bibitem{Carmona:2015haa}
A.~Carmona and M.~Chala, ``{Composite Dark Sectors},''
  \href{http://dx.doi.org/10.1007/JHEP06(2015)105}{{\em JHEP} {\bf 06} (2015)
  105},
\href{http://arxiv.org/abs/1504.00332}{{\tt arXiv:1504.00332 [hep-ph]}}.
%%CITATION = ARXIV:1504.00332;%%.

\bibitem{Lee:2015gsa}
H.~M. Lee and M.-S. Seo, ``{Communication with SIMP dark mesons via Z'
  -portal},'' \href{http://dx.doi.org/10.1016/j.physletb.2015.07.013}{{\em
  Phys. Lett.} {\bf B748} (2015)  316--322},
\href{http://arxiv.org/abs/1504.00745}{{\tt arXiv:1504.00745 [hep-ph]}}.
%%CITATION = ARXIV:1504.00745;%%.

\bibitem{Hochberg:2015vrg}
Y.~Hochberg, E.~Kuflik, and H.~Murayama, ``{SIMP Spectroscopy},''
  \href{http://dx.doi.org/10.1007/JHEP05(2016)090}{{\em JHEP} {\bf 05} (2016)
  090},
\href{http://arxiv.org/abs/1512.07917}{{\tt arXiv:1512.07917 [hep-ph]}}.
%%CITATION = ARXIV:1512.07917;%%.

\bibitem{Bruggisser:2016ixa}
S.~Bruggisser, F.~Riva, and A.~Urbano, ``{Strongly Interacting Light Dark
  Matter},'' \href{http://dx.doi.org/10.21468/SciPostPhys.3.3.017}{{\em SciPost
  Phys.} {\bf 3} (2017) no.~3, 017},
\href{http://arxiv.org/abs/1607.02474}{{\tt arXiv:1607.02474 [hep-ph]}}.
%%CITATION = ARXIV:1607.02474;%%.

\bibitem{Ma:2017vzm}
Y.~Wu, T.~Ma, B.~Zhang, and G.~Cacciapaglia, ``{Composite Dark Matter and
  Higgs},'' \href{http://dx.doi.org/10.1007/JHEP11(2017)058}{{\em JHEP} {\bf
  11} (2017)  058},
\href{http://arxiv.org/abs/1703.06903}{{\tt arXiv:1703.06903 [hep-ph]}}.
%%CITATION = ARXIV:1703.06903;%%.

\bibitem{Davoudiasl:2017zws}
H.~Davoudiasl, P.~P. Giardino, E.~T. Neil, and E.~Rinaldi, ``{Unified Scenario
  for Composite Right-Handed Neutrinos and Dark Matter},''
  \href{http://dx.doi.org/10.1103/PhysRevD.96.115003}{{\em Phys. Rev.} {\bf
  D96} (2017) no.~11, 115003},
\href{http://arxiv.org/abs/1709.01082}{{\tt arXiv:1709.01082 [hep-ph]}}.
%%CITATION = ARXIV:1709.01082;%%.

\bibitem{Berlin:2018tvf}
A.~Berlin, N.~Blinov, S.~Gori, P.~Schuster, and N.~Toro, ``{Cosmology and
  Accelerator Tests of Strongly Interacting Dark Matter},''
  \href{http://dx.doi.org/10.1103/PhysRevD.97.055033}{{\em Phys. Rev.} {\bf
  D97} (2018) no.~5, 055033},
\href{http://arxiv.org/abs/1801.05805}{{\tt arXiv:1801.05805 [hep-ph]}}.
%%CITATION = ARXIV:1801.05805;%%.

\bibitem{Choi:2018iit}
S.-M. Choi, H.~M. Lee, P.~Ko, and A.~Natale, ``{Resolving phenomenological
  problems with strongly-interacting-massive-particle models with dark vector
  resonances},'' \href{http://dx.doi.org/10.1103/PhysRevD.98.015034}{{\em Phys.
  Rev.} {\bf D98} (2018) no.~1, 015034},
\href{http://arxiv.org/abs/1801.07726}{{\tt arXiv:1801.07726 [hep-ph]}}.
%%CITATION = ARXIV:1801.07726;%%.

\bibitem{Hochberg:2018rjs}
Y.~Hochberg, E.~Kuflik, R.~Mcgehee, H.~Murayama, and K.~Schutz, ``{SIMPs
  through the axion portal},''
\href{http://arxiv.org/abs/1806.10139}{{\tt arXiv:1806.10139 [hep-ph]}}.
%%CITATION = ARXIV:1806.10139;%%.

\bibitem{Alves:2009nf}
D.~S. Alves, S.~R. Behbahani, P.~Schuster, and J.~G. Wacker, ``{Composite
  Inelastic Dark Matter},''
  \href{http://dx.doi.org/10.1016/j.physletb.2010.08.006}{{\em Phys.Lett.} {\bf
  B692} (2010)  323--326},
\href{http://arxiv.org/abs/0903.3945}{{\tt arXiv:0903.3945 [hep-ph]}}.
%%CITATION = ARXIV:0903.3945;%%.

\bibitem{Kribs:2009fy}
G.~D. Kribs, T.~S. Roy, J.~Terning, and K.~M. Zurek, ``{Quirky Composite Dark
  Matter},'' \href{http://dx.doi.org/10.1103/PhysRevD.81.095001}{{\em Phys.
  Rev.} {\bf D81} (2010)  095001},
\href{http://arxiv.org/abs/0909.2034}{{\tt arXiv:0909.2034 [hep-ph]}}.
%%CITATION = 0909.2034;%%.

\bibitem{Lisanti:2009am}
M.~Lisanti and J.~G. Wacker, ``{Parity Violation in Composite Inelastic Dark
  Matter Models},'' \href{http://dx.doi.org/10.1103/PhysRevD.82.055023}{{\em
  Phys. Rev.} {\bf D82} (2010)  055023},
\href{http://arxiv.org/abs/0911.4483}{{\tt arXiv:0911.4483 [hep-ph]}}.
%%CITATION = ARXIV:0911.4483;%%.

\bibitem{Alves:2010dd}
D.~Spier Moreira~Alves, S.~R. Behbahani, P.~Schuster, and J.~G. Wacker, ``{The
  Cosmology of Composite Inelastic Dark Matter},''
  \href{http://dx.doi.org/10.1007/JHEP06(2010)113}{{\em JHEP} {\bf 1006} (2010)
   113},
\href{http://arxiv.org/abs/1003.4729}{{\tt arXiv:1003.4729 [hep-ph]}}.
%%CITATION = ARXIV:1003.4729;%%.

\bibitem{Geller:2018biy}
M.~Geller, S.~Iwamoto, G.~Lee, Y.~Shadmi, and O.~Telem, ``{Dark quarkonium
  formation in the early universe},''
  \href{http://dx.doi.org/10.1007/JHEP06(2018)135}{{\em JHEP} {\bf 06} (2018)
  135},
\href{http://arxiv.org/abs/1802.07720}{{\tt arXiv:1802.07720 [hep-ph]}}.
%%CITATION = ARXIV:1802.07720;%%.

\bibitem{Gudnason:2006yj}
S.~B. Gudnason, C.~Kouvaris, and F.~Sannino, ``{Dark Matter from new
  Technicolor Theories},''
  \href{http://dx.doi.org/10.1103/PhysRevD.74.095008}{{\em Phys. Rev.} {\bf
  D74} (2006)  095008},
\href{http://arxiv.org/abs/hep-ph/0608055}{{\tt arXiv:hep-ph/0608055
  [hep-ph]}}.
%%CITATION = HEP-PH/0608055;%%.

\bibitem{Dietrich:2006cm}
D.~D. Dietrich and F.~Sannino, ``{Conformal window of SU(N) gauge theories with
  fermions in higher dimensional representations},''
  \href{http://dx.doi.org/10.1103/PhysRevD.75.085018}{{\em Phys. Rev.} {\bf
  D75} (2007)  085018},
\href{http://arxiv.org/abs/hep-ph/0611341}{{\tt arXiv:hep-ph/0611341
  [hep-ph]}}.
%%CITATION = HEP-PH/0611341;%%.

\bibitem{Foadi:2008qv}
R.~Foadi, M.~T. Frandsen, and F.~Sannino, ``{Technicolor Dark Matter},''
  \href{http://dx.doi.org/10.1103/PhysRevD.80.037702}{{\em Phys. Rev.} {\bf
  D80} (2009)  037702},
\href{http://arxiv.org/abs/0812.3406}{{\tt arXiv:0812.3406 [hep-ph]}}.
%%CITATION = ARXIV:0812.3406;%%.

\bibitem{Mardon:2009gw}
J.~Mardon, Y.~Nomura, and J.~Thaler, ``{Cosmic Signals from the Hidden
  Sector},'' \href{http://dx.doi.org/10.1103/PhysRevD.80.035013}{{\em Phys.
  Rev.} {\bf D80} (2009)  035013},
\href{http://arxiv.org/abs/0905.3749}{{\tt arXiv:0905.3749 [hep-ph]}}.
%%CITATION = ARXIV:0905.3749;%%.

\bibitem{Sannino:2009za}
F.~Sannino, ``{Conformal Dynamics for TeV Physics and Cosmology},'' {\em Acta
  Phys. Polon.} {\bf B40} (2009)  3533--3743,
\href{http://arxiv.org/abs/0911.0931}{{\tt arXiv:0911.0931 [hep-ph]}}.
%%CITATION = ARXIV:0911.0931;%%.

\bibitem{Barbieri:2010mn}
R.~Barbieri, S.~Rychkov, and R.~Torre, ``{Signals of composite
  electroweak-neutral Dark Matter: LHC/Direct Detection interplay},''
  \href{http://dx.doi.org/10.1016/j.physletb.2010.04.010}{{\em Phys. Lett.}
  {\bf B688} (2010)  212--215},
\href{http://arxiv.org/abs/1001.3149}{{\tt arXiv:1001.3149 [hep-ph]}}.
%%CITATION = ARXIV:1001.3149;%%.

\bibitem{Belyaev:2010kp}
A.~Belyaev, M.~T. Frandsen, S.~Sarkar, and F.~Sannino, ``{Mixed dark matter
  from technicolor},'' \href{http://dx.doi.org/10.1103/PhysRevD.83.015007}{{\em
  Phys. Rev.} {\bf D83} (2011)  015007},
\href{http://arxiv.org/abs/1007.4839}{{\tt arXiv:1007.4839 [hep-ph]}}.
%%CITATION = ARXIV:1007.4839;%%.

\bibitem{Appelquist:2013ms}
{\bf Lattice Strong Dynamics (LSD) Collaboration} Collaboration, T.~Appelquist
  {\em et al.}, ``{Lattice calculation of composite dark matter form
  factors},'' \href{http://dx.doi.org/10.1103/PhysRevD.88.014502}{{\em
  Phys.Rev.} {\bf D88} (2013) no.~1, 014502},
\href{http://arxiv.org/abs/1301.1693}{{\tt arXiv:1301.1693 [hep-ph]}}.
%%CITATION = ARXIV:1301.1693;%%.

\bibitem{Hietanen:2013fya}
A.~Hietanen, R.~Lewis, C.~Pica, and F.~Sannino, ``{Composite Goldstone Dark
  Matter: Experimental Predictions from the Lattice},''
\href{http://arxiv.org/abs/1308.4130}{{\tt arXiv:1308.4130 [hep-ph]}}.
%%CITATION = ARXIV:1308.4130;%%.

\bibitem{Cline:2013zca}
J.~M. Cline, Z.~Liu, G.~Moore, and W.~Xue, ``{Composite strongly interacting
  dark matter},'' \href{http://dx.doi.org/10.1103/PhysRevD.90.015023}{{\em
  Phys. Rev.} {\bf D90} (2014) no.~1, 015023},
\href{http://arxiv.org/abs/1312.3325}{{\tt arXiv:1312.3325 [hep-ph]}}.
%%CITATION = ARXIV:1312.3325;%%.

\bibitem{Appelquist:2014jch}
{\bf Lattice Strong Dynamics (LSD) Collaboration} Collaboration, T.~Appelquist
  {\em et al.}, ``{Composite bosonic baryon dark matter on the lattice: SU(4)
  baryon spectrum and the effective Higgs interaction},''
  \href{http://dx.doi.org/10.1103/PhysRevD.89.094508}{{\em Phys.Rev.} {\bf D89}
  (2014) no.~9, 094508},
\href{http://arxiv.org/abs/1402.6656}{{\tt arXiv:1402.6656 [hep-lat]}}.
%%CITATION = ARXIV:1402.6656;%%.

\bibitem{Krnjaic:2014xza}
G.~Krnjaic and K.~Sigurdson, ``{Big Bang Darkleosynthesis},''
  \href{http://dx.doi.org/10.1016/j.physletb.2015.11.001}{{\em Phys. Lett.}
  {\bf B751} (2015)  464--468},
\href{http://arxiv.org/abs/1406.1171}{{\tt arXiv:1406.1171 [hep-ph]}}.
%%CITATION = ARXIV:1406.1171;%%.

\bibitem{Detmold:2014qqa}
W.~Detmold, M.~McCullough, and A.~Pochinsky, ``{Dark Nuclei I: Cosmology and
  Indirect Detection},''
  \href{http://dx.doi.org/10.1103/PhysRevD.90.115013}{{\em Phys.Rev.} {\bf D90}
  (2014) no.~11, 115013},
\href{http://arxiv.org/abs/1406.2276}{{\tt arXiv:1406.2276 [hep-ph]}}.
%%CITATION = ARXIV:1406.2276;%%.

\bibitem{Detmold:2014kba}
W.~Detmold, M.~McCullough, and A.~Pochinsky, ``{Dark nuclei. II. Nuclear
  spectroscopy in two-color QCD},''
  \href{http://dx.doi.org/10.1103/PhysRevD.90.114506}{{\em Phys.Rev.} {\bf D90}
  (2014) no.~11, 114506},
\href{http://arxiv.org/abs/1406.4116}{{\tt arXiv:1406.4116 [hep-lat]}}.
%%CITATION = ARXIV:1406.4116;%%.

\bibitem{Brod:2014loa}
J.~Brod, J.~Drobnak, A.~L. Kagan, E.~Stamou, and J.~Zupan, ``{Stealth QCD-like
  strong interactions and the $t \bar {t}$ asymmetry},''
  \href{http://dx.doi.org/10.1103/PhysRevD.91.095009}{{\em Phys. Rev.} {\bf
  D91} (2015) no.~9, 095009},
\href{http://arxiv.org/abs/1407.8188}{{\tt arXiv:1407.8188 [hep-ph]}}.
%%CITATION = ARXIV:1407.8188;%%.

\bibitem{Asano:2014wra}
M.~Asano and R.~Kitano, ``{Partially Composite Dark Matter},''
  \href{http://dx.doi.org/10.1007/JHEP09(2014)171}{{\em JHEP} {\bf 09} (2014)
  171},
\href{http://arxiv.org/abs/1406.6374}{{\tt arXiv:1406.6374 [hep-ph]}}.
%%CITATION = ARXIV:1406.6374;%%.

\bibitem{Appelquist:2015yfa}
T.~Appelquist {\em et al.}, ``{Stealth Dark Matter: Dark scalar baryons through
  the Higgs portal},'' \href{http://dx.doi.org/10.1103/PhysRevD.92.075030}{{\em
  Phys. Rev.} {\bf D92} (2015) no.~7, 075030},
\href{http://arxiv.org/abs/1503.04203}{{\tt arXiv:1503.04203 [hep-ph]}}.
%%CITATION = ARXIV:1503.04203;%%.

\bibitem{Appelquist:2015zfa}
T.~Appelquist {\em et al.}, ``{Detecting Stealth Dark Matter Directly through
  Electromagnetic Polarizability},''
  \href{http://dx.doi.org/10.1103/PhysRevLett.115.171803}{{\em Phys. Rev.
  Lett.} {\bf 115} (2015) no.~17, 171803},
\href{http://arxiv.org/abs/1503.04205}{{\tt arXiv:1503.04205 [hep-ph]}}.
%%CITATION = ARXIV:1503.04205;%%.

\bibitem{Drach:2015epq}
V.~Drach, A.~Hietanen, C.~Pica, J.~Rantaharju, and F.~Sannino, ``{Template
  Composite Dark Matter: $SU(2)$ gauge theory with 2 fundamental flavours},''
  \href{http://dx.doi.org/10.22323/1.251.0234}{{\em PoS} {\bf LATTICE2015}
  (2016)  234},
\href{http://arxiv.org/abs/1511.04370}{{\tt arXiv:1511.04370 [hep-lat]}}.
%%CITATION = ARXIV:1511.04370;%%.

\bibitem{Fichet:2016clq}
S.~Fichet, ``{Shining Light on Polarizable Dark Particles},''
  \href{http://dx.doi.org/10.1007/JHEP04(2017)088}{{\em JHEP} {\bf 04} (2017)
  088},
\href{http://arxiv.org/abs/1609.01762}{{\tt arXiv:1609.01762 [hep-ph]}}.
%%CITATION = ARXIV:1609.01762;%%.

\bibitem{Co:2016akw}
R.~T. Co, K.~Harigaya, and Y.~Nomura, ``{Chiral Dark Sector},''
  \href{http://dx.doi.org/10.1103/PhysRevLett.118.101801}{{\em Phys. Rev.
  Lett.} {\bf 118} (2017) no.~10, 101801},
\href{http://arxiv.org/abs/1610.03848}{{\tt arXiv:1610.03848 [hep-ph]}}.
%%CITATION = ARXIV:1610.03848;%%.

\bibitem{Dienes:2016vei}
K.~R. Dienes, F.~Huang, S.~Su, and B.~Thomas, ``{Dynamical Dark Matter from
  Strongly-Coupled Dark Sectors},''
  \href{http://dx.doi.org/10.1103/PhysRevD.95.043526}{{\em Phys. Rev.} {\bf
  D95} (2017) no.~4, 043526},
\href{http://arxiv.org/abs/1610.04112}{{\tt arXiv:1610.04112 [hep-ph]}}.
%%CITATION = ARXIV:1610.04112;%%.

\bibitem{Ishida:2016fbp}
H.~Ishida, S.~Matsuzaki, and Y.~Yamaguchi, ``{Bosonic-Seesaw Portal Dark
  Matter},'' \href{http://dx.doi.org/10.1093/ptep/ptx132}{{\em PTEP} {\bf 2017}
  (2017) no.~10, 103B01},
\href{http://arxiv.org/abs/1610.07137}{{\tt arXiv:1610.07137 [hep-ph]}}.
%%CITATION = ARXIV:1610.07137;%%.

\bibitem{Francis:2016bzf}
A.~Francis, R.~J. Hudspith, R.~Lewis, and S.~Tulin, ``{Dark matter from
  one-flavor SU(2) gauge theory},''
  \href{http://dx.doi.org/10.22323/1.256.0227}{{\em PoS} {\bf LATTICE2016}
  (2016)  227},
\href{http://arxiv.org/abs/1610.10068}{{\tt arXiv:1610.10068 [hep-lat]}}.
%%CITATION = ARXIV:1610.10068;%%.

\bibitem{Lonsdale:2017mzg}
S.~J. Lonsdale, M.~Schroor, and R.~R. Volkas, ``{Asymmetric Dark Matter and the
  hadronic spectra of hidden QCD},''
  \href{http://dx.doi.org/10.1103/PhysRevD.96.055027}{{\em Phys. Rev.} {\bf
  D96} (2017) no.~5, 055027},
\href{http://arxiv.org/abs/1704.05213}{{\tt arXiv:1704.05213 [hep-ph]}}.
%%CITATION = ARXIV:1704.05213;%%.

\bibitem{Berryman:2017twh}
J.~M. Berryman, A.~de~Gouvêa, K.~J. Kelly, and Y.~Zhang, ``{Dark Matter and
  Neutrino Mass from the Smallest Non-Abelian Chiral Dark Sector},''
  \href{http://dx.doi.org/10.1103/PhysRevD.96.075010}{{\em Phys. Rev.} {\bf
  D96} (2017) no.~7, 075010},
\href{http://arxiv.org/abs/1706.02722}{{\tt arXiv:1706.02722 [hep-ph]}}.
%%CITATION = ARXIV:1706.02722;%%.

\bibitem{Mitridate:2017oky}
A.~Mitridate, M.~Redi, J.~Smirnov, and A.~Strumia, ``{Dark Matter as a weakly
  coupled Dark Baryon},'' \href{http://dx.doi.org/10.1007/JHEP10(2017)210}{{\em
  JHEP} {\bf 10} (2017)  210},
\href{http://arxiv.org/abs/1707.05380}{{\tt arXiv:1707.05380 [hep-ph]}}.
%%CITATION = ARXIV:1707.05380;%%.

\bibitem{Francis:2018xjd}
A.~Francis, R.~J. Hudspith, R.~Lewis, and S.~Tulin, ``{Dark Matter from Strong
  Dynamics: The Minimal Theory of Dark Baryons},''
\href{http://arxiv.org/abs/1809.09117}{{\tt arXiv:1809.09117 [hep-ph]}}.
%%CITATION = ARXIV:1809.09117;%%.

\bibitem{Kribs:2016cew}
G.~D. Kribs and E.~T. Neil, ``{Review of strongly-coupled composite dark matter
  models and lattice simulations},''
  \href{http://dx.doi.org/10.1142/S0217751X16430041}{{\em Int. J. Mod. Phys.}
  {\bf A31} (2016) no.~22, 1643004},
\href{http://arxiv.org/abs/1604.04627}{{\tt arXiv:1604.04627 [hep-ph]}}.
%%CITATION = ARXIV:1604.04627;%%.

\bibitem{Kilic:2009mi}
C.~Kilic, T.~Okui, and R.~Sundrum, ``{Vectorlike Confinement at the LHC},''
  \href{http://dx.doi.org/10.1007/JHEP02(2010)018}{{\em JHEP} {\bf 02} (2010)
  018},
\href{http://arxiv.org/abs/0906.0577}{{\tt arXiv:0906.0577 [hep-ph]}}.
%%CITATION = ARXIV:0906.0577;%%.

\bibitem{Kilic:2010et}
C.~Kilic and T.~Okui, ``{The LHC Phenomenology of Vectorlike Confinement},''
  \href{http://dx.doi.org/10.1007/JHEP04(2010)128}{{\em JHEP} {\bf 04} (2010)
  128},
\href{http://arxiv.org/abs/1001.4526}{{\tt arXiv:1001.4526 [hep-ph]}}.
%%CITATION = ARXIV:1001.4526;%%.

\bibitem{Harnik:2011mv}
R.~Harnik, G.~D. Kribs, and A.~Martin, ``{Quirks at the Tevatron and Beyond},''
  \href{http://dx.doi.org/10.1103/PhysRevD.84.035029}{{\em Phys. Rev.} {\bf
  D84} (2011)  035029},
\href{http://arxiv.org/abs/1106.2569}{{\tt arXiv:1106.2569 [hep-ph]}}.
%%CITATION = ARXIV:1106.2569;%%.

\bibitem{Fok:2011yc}
R.~Fok and G.~D. Kribs, ``{Chiral Quirkonium Decays},''
  \href{http://dx.doi.org/10.1103/PhysRevD.84.035001}{{\em Phys. Rev.} {\bf
  D84} (2011)  035001},
\href{http://arxiv.org/abs/1106.3101}{{\tt arXiv:1106.3101 [hep-ph]}}.
%%CITATION = ARXIV:1106.3101;%%.

\bibitem{Bai:2013xga}
Y.~Bai and P.~Schwaller, ``{Scale of dark QCD},''
  \href{http://dx.doi.org/10.1103/PhysRevD.89.063522}{{\em Phys. Rev.} {\bf
  D89} (2014) no.~6, 063522},
\href{http://arxiv.org/abs/1306.4676}{{\tt arXiv:1306.4676 [hep-ph]}}.
%%CITATION = ARXIV:1306.4676;%%.

\bibitem{Chacko:2015fbc}
Z.~Chacko, D.~Curtin, and C.~B. Verhaaren, ``{A Quirky Probe of Neutral
  Naturalness},'' \href{http://dx.doi.org/10.1103/PhysRevD.94.011504}{{\em
  Phys. Rev.} {\bf D94} (2016) no.~1, 011504},
\href{http://arxiv.org/abs/1512.05782}{{\tt arXiv:1512.05782 [hep-ph]}}.
%%CITATION = ARXIV:1512.05782;%%.

\bibitem{Agashe:2016rle}
K.~Agashe, P.~Du, S.~Hong, and R.~Sundrum, ``{Flavor Universal Resonances and
  Warped Gravity},'' \href{http://dx.doi.org/10.1007/JHEP01(2017)016}{{\em
  JHEP} {\bf 01} (2017)  016},
\href{http://arxiv.org/abs/1608.00526}{{\tt arXiv:1608.00526 [hep-ph]}}.
%%CITATION = ARXIV:1608.00526;%%.

\bibitem{Matsuzaki:2017bpp}
S.~Matsuzaki, K.~Nishiwaki, and R.~Watanabe, ``{Phenomenology of flavorful
  composite vector bosons in light of $B$ anomalies},''
  \href{http://dx.doi.org/10.1007/JHEP08(2017)145}{{\em JHEP} {\bf 08} (2017)
  145},
\href{http://arxiv.org/abs/1706.01463}{{\tt arXiv:1706.01463 [hep-ph]}}.
%%CITATION = ARXIV:1706.01463;%%.

\bibitem{Draper:2018tmh}
P.~Draper, J.~Kozaczuk, and J.-H. Yu, ``{Theta in new QCD-like sectors},''
  \href{http://dx.doi.org/10.1103/PhysRevD.98.015028}{{\em Phys. Rev.} {\bf
  D98} (2018) no.~1, 015028},
\href{http://arxiv.org/abs/1803.00015}{{\tt arXiv:1803.00015 [hep-ph]}}.
%%CITATION = ARXIV:1803.00015;%%.

\bibitem{Buttazzo:2018qqp}
D.~Buttazzo, D.~Redigolo, F.~Sala, and A.~Tesi, ``{Fusing Vectors into Scalars
  at High Energy Lepton Colliders},''
\href{http://arxiv.org/abs/1807.04743}{{\tt arXiv:1807.04743 [hep-ph]}}.
%%CITATION = ARXIV:1807.04743;%%.

\bibitem{Schwaller:2015gea}
P.~Schwaller, D.~Stolarski, and A.~Weiler, ``{Emerging Jets},''
  \href{http://dx.doi.org/10.1007/JHEP05(2015)059}{{\em JHEP} {\bf 05} (2015)
  059},
\href{http://arxiv.org/abs/1502.05409}{{\tt arXiv:1502.05409 [hep-ph]}}.
%%CITATION = ARXIV:1502.05409;%%.

\bibitem{Cohen:2015toa}
T.~Cohen, M.~Lisanti, and H.~K. Lou, ``{Semivisible Jets: Dark Matter
  Undercover at the LHC},''
  \href{http://dx.doi.org/10.1103/PhysRevLett.115.171804}{{\em Phys. Rev.
  Lett.} {\bf 115} (2015) no.~17, 171804},
\href{http://arxiv.org/abs/1503.00009}{{\tt arXiv:1503.00009 [hep-ph]}}.
%%CITATION = ARXIV:1503.00009;%%.

\bibitem{Freytsis:2016dgf}
M.~Freytsis, S.~Knapen, D.~J. Robinson, and Y.~Tsai, ``{Gamma-rays from Dark
  Showers with Twin Higgs Models},''
  \href{http://dx.doi.org/10.1007/JHEP05(2016)018}{{\em JHEP} {\bf 05} (2016)
  018},
\href{http://arxiv.org/abs/1601.07556}{{\tt arXiv:1601.07556 [hep-ph]}}.
%%CITATION = ARXIV:1601.07556;%%.

\bibitem{Zhang:2016sll}
M.~Kim, H.-S. Lee, M.~Park, and M.~Zhang, ``{Examining the origin of dark
  matter mass at colliders},''
\href{http://arxiv.org/abs/1612.02850}{{\tt arXiv:1612.02850 [hep-ph]}}.
%%CITATION = ARXIV:1612.02850;%%.

\bibitem{Cohen:2017pzm}
T.~Cohen, M.~Lisanti, H.~K. Lou, and S.~Mishra-Sharma, ``{LHC Searches for Dark
  Sector Showers},'' \href{http://dx.doi.org/10.1007/JHEP11(2017)196}{{\em
  JHEP} {\bf 11} (2017)  196},
\href{http://arxiv.org/abs/1707.05326}{{\tt arXiv:1707.05326 [hep-ph]}}.
%%CITATION = ARXIV:1707.05326;%%.

\bibitem{Beauchesne:2017yhh}
H.~Beauchesne, E.~Bertuzzo, G.~Grilli Di~Cortona, and Z.~Tabrizi, ``{Collider
  phenomenology of Hidden Valley mediators of spin 0 or 1/2 with semivisible
  jets},'' \href{http://dx.doi.org/10.1007/JHEP08(2018)030}{{\em JHEP} {\bf 08}
  (2018)  030},
\href{http://arxiv.org/abs/1712.07160}{{\tt arXiv:1712.07160 [hep-ph]}}.
%%CITATION = ARXIV:1712.07160;%%.

\bibitem{Renner:2018fhh}
S.~Renner and P.~Schwaller, ``{A flavoured dark sector},''
  \href{http://dx.doi.org/10.1007/JHEP08(2018)052}{{\em JHEP} {\bf 08} (2018)
  052},
\href{http://arxiv.org/abs/1803.08080}{{\tt arXiv:1803.08080 [hep-ph]}}.
%%CITATION = ARXIV:1803.08080;%%.

\bibitem{Mahbubani:2017gjh}
R.~Mahbubani, P.~Schwaller, and J.~Zurita, ``{Closing the window for compressed
  Dark Sectors with disappearing charged tracks},''
  \href{http://dx.doi.org/10.1007/JHEP06(2017)119,
  10.1007/JHEP10(2017)061}{{\em JHEP} {\bf 06} (2017)  119},
  \href{http://arxiv.org/abs/1703.05327}{{\tt arXiv:1703.05327 [hep-ph]}}.
[Erratum: JHEP10,061(2017)].
%%CITATION = ARXIV:1703.05327;%%.

\bibitem{Buchmueller:2017uqu}
O.~Buchmueller, A.~De~Roeck, K.~Hahn, M.~McCullough, P.~Schwaller, K.~Sung, and
  T.-T. Yu, ``{Simplified Models for Displaced Dark Matter Signatures},''
  \href{http://dx.doi.org/10.1007/JHEP09(2017)076}{{\em JHEP} {\bf 09} (2017)
  076},
\href{http://arxiv.org/abs/1704.06515}{{\tt arXiv:1704.06515 [hep-ph]}}.
%%CITATION = ARXIV:1704.06515;%%.

\bibitem{Daci:2015hca}
N.~Daci, I.~De~Bruyn, S.~Lowette, M.~H.~G. Tytgat, and B.~Zaldivar,
  ``{Simplified SIMPs and the LHC},''
  \href{http://dx.doi.org/10.1007/JHEP11(2015)108}{{\em JHEP} {\bf 11} (2015)
  108},
\href{http://arxiv.org/abs/1503.05505}{{\tt arXiv:1503.05505 [hep-ph]}}.
%%CITATION = ARXIV:1503.05505;%%.

\bibitem{Hochberg:2017khi}
Y.~Hochberg, E.~Kuflik, and H.~Murayama, ``{Dark spectroscopy at lepton
  colliders},'' \href{http://dx.doi.org/10.1103/PhysRevD.97.055030}{{\em Phys.
  Rev.} {\bf D97} (2018) no.~5, 055030},
\href{http://arxiv.org/abs/1706.05008}{{\tt arXiv:1706.05008 [hep-ph]}}.
%%CITATION = ARXIV:1706.05008;%%.

\bibitem{Han:2007ae}
T.~Han, Z.~Si, K.~M. Zurek, and M.~J. Strassler, ``{Phenomenology of hidden
  valleys at hadron colliders},''
  \href{http://dx.doi.org/10.1088/1126-6708/2008/07/008}{{\em JHEP} {\bf 07}
  (2008)  008},
\href{http://arxiv.org/abs/0712.2041}{{\tt arXiv:0712.2041 [hep-ph]}}.
%%CITATION = ARXIV:0712.2041;%%.

\bibitem{Kang:2008ea}
J.~Kang and M.~A. Luty, ``{Macroscopic Strings and 'Quirks' at Colliders},''
  \href{http://dx.doi.org/10.1088/1126-6708/2009/11/065}{{\em JHEP} {\bf 11}
  (2009)  065},
\href{http://arxiv.org/abs/0805.4642}{{\tt arXiv:0805.4642 [hep-ph]}}.
%%CITATION = ARXIV:0805.4642;%%.

\bibitem{Harnik:2008ax}
R.~Harnik and T.~Wizansky, ``{Signals of New Physics in the Underlying
  Event},'' \href{http://dx.doi.org/10.1103/PhysRevD.80.075015}{{\em Phys.
  Rev.} {\bf D80} (2009)  075015},
\href{http://arxiv.org/abs/0810.3948}{{\tt arXiv:0810.3948 [hep-ph]}}.
%%CITATION = ARXIV:0810.3948;%%.

\bibitem{Knapen:2016hky}
S.~Knapen, S.~Pagan~Griso, M.~Papucci, and D.~J. Robinson, ``{Triggering Soft
  Bombs at the LHC},'' \href{http://dx.doi.org/10.1007/JHEP08(2017)076}{{\em
  JHEP} {\bf 08} (2017)  076},
\href{http://arxiv.org/abs/1612.00850}{{\tt arXiv:1612.00850 [hep-ph]}}.
%%CITATION = ARXIV:1612.00850;%%.

\bibitem{Pierce:2017taw}
A.~Pierce, B.~Shakya, Y.~Tsai, and Y.~Zhao, ``{Searching for confining hidden
  valleys at LHCb, ATLAS, and CMS},''
  \href{http://dx.doi.org/10.1103/PhysRevD.97.095033}{{\em Phys. Rev.} {\bf
  D97} (2018) no.~9, 095033},
\href{http://arxiv.org/abs/1708.05389}{{\tt arXiv:1708.05389 [hep-ph]}}.
%%CITATION = ARXIV:1708.05389;%%.

\bibitem{Beylin:2016kga}
V.~Beylin, M.~Bezuglov, V.~Kuksa, and N.~Volchanskiy, ``{An analysis of a
  minimal vectorlike extension of the Standard Model},''
  \href{http://dx.doi.org/10.1155/2017/1765340}{{\em Adv. High Energy Phys.}
  {\bf 2017} (2017)  1765340},
\href{http://arxiv.org/abs/1611.06006}{{\tt arXiv:1611.06006 [hep-ph]}}.
%%CITATION = ARXIV:1611.06006;%%.

\bibitem{Kribs:2018ilo}
G.~D. Kribs, A.~Martin, B.~Ostdiek, and T.~Tong, ``{Dark Mesons at the LHC},''
\href{http://arxiv.org/abs/1809.10184}{{\tt arXiv:1809.10184 [hep-ph]}}.
%%CITATION = ARXIV:1809.10184;%%.

\bibitem{Bai:2010mn}
Y.~Bai and A.~Martin, ``{Topological Pions},''
  \href{http://dx.doi.org/10.1016/j.physletb.2010.08.058}{{\em Phys. Lett.}
  {\bf B693} (2010)  292--295},
\href{http://arxiv.org/abs/1003.3006}{{\tt arXiv:1003.3006 [hep-ph]}}.
%%CITATION = ARXIV:1003.3006;%%.

\bibitem{Duka:1999uc}
P.~Duka, J.~Gluza, and M.~Zralek, ``{Quantization and renormalization of the
  manifest left-right symmetric model of electroweak interactions},''
  \href{http://dx.doi.org/10.1006/aphy.1999.5988}{{\em Annals Phys.} {\bf 280}
  (2000)  336--408},
\href{http://arxiv.org/abs/hep-ph/9910279}{{\tt arXiv:hep-ph/9910279
  [hep-ph]}}.
%%CITATION = HEP-PH/9910279;%%.

\bibitem{Buchalla:2014eca}
G.~Buchalla, O.~Cata, and C.~Krause, ``{A Systematic Approach to the SILH
  Lagrangian},'' \href{http://dx.doi.org/10.1016/j.nuclphysb.2015.03.024}{{\em
  Nucl. Phys.} {\bf B894} (2015)  602--620},
\href{http://arxiv.org/abs/1412.6356}{{\tt arXiv:1412.6356 [hep-ph]}}.
%%CITATION = ARXIV:1412.6356;%%.

\bibitem{Craig:2012pu}
N.~Craig, J.~A. Evans, R.~Gray, C.~Kilic, M.~Park, S.~Somalwar, and S.~Thomas,
  ``{Multi-Lepton Signals of Multiple Higgs Bosons},''
  \href{http://dx.doi.org/10.1007/JHEP02(2013)033}{{\em JHEP} {\bf 02} (2013)
  033},
\href{http://arxiv.org/abs/1210.0559}{{\tt arXiv:1210.0559 [hep-ph]}}.
%%CITATION = ARXIV:1210.0559;%%.

\bibitem{Gunion:2002zf}
J.~F. Gunion and H.~E. Haber, ``{The CP conserving two Higgs doublet model: The
  Approach to the decoupling limit},''
  \href{http://dx.doi.org/10.1103/PhysRevD.67.075019}{{\em Phys. Rev.} {\bf
  D67} (2003)  075019},
\href{http://arxiv.org/abs/hep-ph/0207010}{{\tt arXiv:hep-ph/0207010
  [hep-ph]}}.
%%CITATION = HEP-PH/0207010;%%.

\bibitem{Djouadi:2005gj}
A.~Djouadi, ``{The Anatomy of electro-weak symmetry breaking. II. The Higgs
  bosons in the minimal supersymmetric model},''
  \href{http://dx.doi.org/10.1016/j.physrep.2007.10.005}{{\em Phys. Rept.} {\bf
  459} (2008)  1--241},
\href{http://arxiv.org/abs/hep-ph/0503173}{{\tt arXiv:hep-ph/0503173
  [hep-ph]}}.
%%CITATION = HEP-PH/0503173;%%.

\bibitem{Das:1967it}
T.~Das, G.~S. Guralnik, V.~S. Mathur, F.~E. Low, and J.~E. Young,
  ``{Electromagnetic mass difference of pions},''
\href{http://dx.doi.org/10.1103/PhysRevLett.18.759}{{\em Phys. Rev. Lett.} {\bf
  18} (1967)  759--761}.
%%CITATION = PRLTA,18,759;%%.

\bibitem{Callan:1969sn}
C.~G. Callan, Jr., S.~R. Coleman, J.~Wess, and B.~Zumino, ``{Structure of
  phenomenological Lagrangians. 2.},''
\href{http://dx.doi.org/10.1103/PhysRev.177.2247}{{\em Phys. Rev.} {\bf 177}
  (1969)  2247--2250}.
%%CITATION = PHRVA,177,2247;%%.

\bibitem{Bando:1987br}
M.~Bando, T.~Kugo, and K.~Yamawaki, ``{Nonlinear Realization and Hidden Local
  Symmetries},''
\href{http://dx.doi.org/10.1016/0370-1573(88)90019-1}{{\em Phys. Rept.} {\bf
  164} (1988)  217--314}.
%%CITATION = PRPLC,164,217;%%.

\bibitem{SekharChivukula:2007gi}
R.~S. Chivukula, N.~D. Christensen, and E.~H. Simmons, ``{Low-energy effective
  theory, unitarity, and non-decoupling behavior in a model with heavy
  Higgs-triplet fields},''
  \href{http://dx.doi.org/10.1103/PhysRevD.77.035001}{{\em Phys. Rev.} {\bf
  D77} (2008)  035001},
\href{http://arxiv.org/abs/0712.0546}{{\tt arXiv:0712.0546 [hep-ph]}}.
%%CITATION = ARXIV:0712.0546;%%.

\bibitem{Gunion:1989we}
J.~F. Gunion, H.~E. Haber, G.~L. Kane, and S.~Dawson, ``{The Higgs Hunter's
  Guide},''
{\em Front. Phys.} {\bf 80} (2000)  1--404.
%%CITATION = FRPHA,80,1;%%.

\bibitem{usgm}
G.~D. Kribs, A.~Martin, and T.~Tong, ``work in progress,''.

\bibitem{Nagel:2004sw}
F.~Nagel, {\em {New aspects of gauge-boson couplings and the Higgs sector}}.
\newblock PhD thesis, Heidelberg U., 2004.
\newblock
\url{http://www.ub.uni-heidelberg.de/archiv/4803}.
\newblock
%%CITATION = INSPIRE-672849;%%.

\bibitem{Maniatis:2006fs}
M.~Maniatis, A.~von Manteuffel, O.~Nachtmann, and F.~Nagel, ``{Stability and
  symmetry breaking in the general two-Higgs-doublet model},''
  \href{http://dx.doi.org/10.1140/epjc/s10052-006-0016-6}{{\em Eur. Phys. J.}
  {\bf C48} (2006)  805--823},
\href{http://arxiv.org/abs/hep-ph/0605184}{{\tt arXiv:hep-ph/0605184
  [hep-ph]}}.
%%CITATION = HEP-PH/0605184;%%.

\bibitem{Grzadkowski:2010dj}
B.~Grzadkowski, M.~Maniatis, and J.~Wudka, ``{The bilinear formalism and the
  custodial symmetry in the two-Higgs-doublet model},''
  \href{http://dx.doi.org/10.1007/JHEP11(2011)030}{{\em JHEP} {\bf 11} (2011)
  030},
\href{http://arxiv.org/abs/1011.5228}{{\tt arXiv:1011.5228 [hep-ph]}}.
%%CITATION = ARXIV:1011.5228;%%.

\bibitem{Haber:2010bw}
H.~E. Haber and D.~O'Neil, ``{Basis-independent methods for the
  two-Higgs-doublet model III: The CP-conserving limit, custodial symmetry, and
  the oblique parameters S, T, U},''
  \href{http://dx.doi.org/10.1103/PhysRevD.83.055017}{{\em Phys. Rev.} {\bf
  D83} (2011)  055017},
\href{http://arxiv.org/abs/1011.6188}{{\tt arXiv:1011.6188 [hep-ph]}}.
%%CITATION = ARXIV:1011.6188;%%.

\bibitem{Haber:1993an}
H.~E. Haber and R.~Hempfling, ``{The Renormalization group improved Higgs
  sector of the minimal supersymmetric model},''
  \href{http://dx.doi.org/10.1103/PhysRevD.48.4280}{{\em Phys. Rev.} {\bf D48}
  (1993)  4280--4309},
\href{http://arxiv.org/abs/hep-ph/9307201}{{\tt arXiv:hep-ph/9307201
  [hep-ph]}}.
%%CITATION = HEP-PH/9307201;%%.

\bibitem{Pomarol:1993mu}
A.~Pomarol and R.~Vega, ``{Constraints on CP violation in the Higgs sector from
  the rho parameter},''
  \href{http://dx.doi.org/10.1016/0550-3213(94)90611-4}{{\em Nucl. Phys.} {\bf
  B413} (1994)  3--15},
\href{http://arxiv.org/abs/hep-ph/9305272}{{\tt arXiv:hep-ph/9305272
  [hep-ph]}}.
%%CITATION = HEP-PH/9305272;%%.

\end{thebibliography}\endgroup

\end{document}